\documentclass[12pt]{emulateapj}
\usepackage{natbib}
\usepackage{color}
\usepackage{hyperref} %----set up hyper link to citations, equations...
\hypersetup{colorlinks=true,citecolor=blue}
\def \bea {\begin{eqnarray}}
\def \ena {\end{eqnarray}}               
\def \bee {\begin{equation}}
\def \ene {\end{equation}}
\def    \simlt  {\lower.5ex\hbox{$\; \buildrel < \over \sim \;$}}
\def    \simgt  {\lower.5ex\hbox{$\; \buildrel > \over \sim \;$}}

\newcommand     \mum    {\,\mu{\rm m}}  % to use in math mode

\def	\cm		{\,{\rm {cm}}}

\def	\erg		{\,{\rm {erg}}}
\def	\eV		{\,{\rm {eV}}\,}
\def    \exp 		{\,{\rm {exp}}}
\def	\g		{\,{\rm g}}

\def	\K		{\,{\rm K}}

\def	\AU		{\,{\rm {AU}}}
\def	\pc		{\,{\rm {pc}}}

\def	\s		{\,{\rm s}}

\def    \yr  		{\,{\rm {yr}}}

\def	\H		{\rm H}

\def    \Bv     	{\bf  B}

\def	\rad		{\rm {rad}}

\def    \abs     	{\rm {abs}}

\def    \coll        	{\rm {coll}}

\def	\coll		{\rm {coll}}

\begin{document}
\shorttitle{Damage to spacecraft}
\shortauthors{Hoang, Lazarian, Burkhart, \& Loeb}
\title{The Interaction of relativistic spacecrafts with the interstellar medium}
\author{Thiem Hoang\altaffilmark{1,2}, A. Lazarian\altaffilmark{3}, Blakesley Burkhart\altaffilmark{4}, and Abraham Loeb\altaffilmark{4}}

\altaffiltext{1}{Korea Astronomy and Space Science Institute 776, Daedeokdae-ro, Yuseong-gu, Daejeon 34055, Korea;\href{mailto:thiemhoang@kasi.re.kr}{thiemhoang@kasi.re.kr}}
\altaffiltext{2}{Canadian Institute for Theoretical Astrophysics, University of Toronto, 60 St. George Street, Toronto, ON M5S 3H8, Canada}
\altaffiltext{3}{Department of Astronomy, University of Wisconsin-Madison, Madison, WI 53705, USA}
\altaffiltext{4}{Harvard-Smithsonian Center for Astrophysics, 60 Garden st.,  Cambridge, MA, USA}

\begin{abstract}
The Breakthrough Starshot initiative aims to launch {a gram-scale spacecraft to a speed of $v\sim 0.2$c, capable of reaching the nearest star system, $\alpha$ Centauri, in about 20 years.} However, a critical challenge for the initiative is the damage to the spacecraft by interstellar gas and dust during the journey. In this paper, we quantify the interaction of a relativistic spacecraft with gas and dust in the interstellar medium. For gas bombardment, we find that damage by track formation due to heavy elements is an important effect. We find that gas bombardment can potentially damage the surface of the spacecraft to a depth of $\sim 0.1$ mm { for quartz material} after traversing a gas column of $N_{\H}\sim 2\times 10^{18}\cm^{-2}$ { along the path to $\alpha$ Centauri}, whereas the effect is much weaker for graphite material. The effect of dust bombardment erodes the spacecraft surface and produces numerous craters due to explosive evaporation of surface atoms. For a spacecraft speed $v=0.2c$, we find that dust bombardment can erode a surface layer of $\sim 0.5$ mm thickness after the spacecraft has swept a column density of $N_{\H}\sim 3\times 10^{17}\cm^{-2}$, assuming the standard gas-to-dust ratio of the interstellar medium. Dust bombardment also damages the spacecraft surface by modifying the material structure through melting. We calculate the equilibrium surface temperature due to collisional heating by gas atoms as well as the temperature profile as a function of depth into the spacecraft. Our quantitative results suggest methods for damage control, and we highlight possibilities for shielding strategies and protection of the spacecraft. 

\keywords{interstellar medium, interplanetary medium, spacecraft}
\end{abstract}

\section{Introduction}\label{sec:intro}
The Breakthrough Starshot initiative\footnote{https://breakthroughinitiatives.org/Initiative/3} aims to launch gram-scale spacecrafts { with miniaturized electronic components (such as camera, navigation, and communication systems)} to relativistic speeds ($v\sim 0.2c)$. This will enable the spacecraft to reach the nearest stars, like $\alpha$ Centauri (distance of 1.34 pc), within a human lifetime. Such spacecrafts would also revolutionize human exploration of the solar system, the neighboring Oort cloud, and the local interstellar medium (ISM). Given the potential feasibility of the suggested technology \footnote{Breakthrough Starshot involves a phased laser array propulsion system to accelerate a gram-scale reflective sail and electronic instruments.  We highlight the proposed technological set-up in Section 2.1.} to accelerate a small spacecraft to relativistic speeds, the next essential question concerns the effects of the interplanetary and interstellar media on the spacecrafts. Will a spacecraft moving relativistically be able to sustain the damage inflicted by the interstellar gas and dust?

On its journey through the interstellar medium, a relativistic spacecraft will collide with interstellar atoms and dust grains. In the rest frame of the spacecraft, {the external atoms will stream relativistically and their bombardment will damage the surface of the spacecraft and pose a potential challenge for its sensitive electronic components.} A quantitative study of interstellar gas and dust interactions with the spacecraft is necessary for engineering a system that is able to protect the spacecraft.

This paper makes use of previous studies on the destruction of fast moving dust grains. In particular, \cite*{2015ApJ...806..255H} (hereafter HLS15) studied the destruction of relativistic grains in various environmental conditions by a number of physical processes, including thermal sublimation, electronic sputtering, grain-grain collisions, and Coulomb explosions.  HLS15 identified that, for relativistic dust, the most important damage processes are Coulomb explosion and explosive evaporation following grain-grain collisions. In light of this study, we might also expect that damage of $v\sim 0.1c$ spacecraft is dominated by collisions with ambient dust grains. {The rate of hits by interstellar dust was estimated by \cite{2016arXiv160401356L} for a variety of shapes of the spacecraft. Our paper evaluates the damage both by interstellar dust grains and gas atoms through detailed treatment of interactions.}

The structure of the paper is as follows. We first present a general description of the Breakthrough Starshot program and spacecraft properties in Section \ref{sec:des}. Then, we discuss the general physics involved for the relativistic spacecraft's interaction with interstellar gas in Section \ref{sec:physics}, and quantify the damage to the spacecraft by gas bombardment in Section \ref{sec:ISMgas}. Section \ref{sec:ISMdust} is devoted to the interaction of dust grains with the spacecraft. The problem of heating due to collisions and radiation field is treated in Section \ref{sec:heat}. We present an extended discussion in Section \ref{sec:dis} and summarize our results in Section \ref{sec:summ}.

\section{General description of spacecrafts and model parameters}\label{sec:des}

Breakthrough Starshot\footnote{https://breakthroughinitiatives.org/Concept/3} is a research and engineering program aiming to demonstrate proof-of-concept for new technology enabling ultra-light unmanned space flight at 20\% of the speed of light and to lay the foundations for a flyby mission to $\alpha$ Centauri within a generation. 
The spacecraft design is expected to consist of two main components: a Starchip and the lightsail. The lightsail is expected to be {less than $1\mum$ thick} and made of a highly reflective material such as graphene-based materials. The Starchip contains electronic instruments (i.e. sensors, cameras, etc.), presumably made of semiconductor material such as quartz. Therefore, hereafter, we discuss the effects of interstellar matter on a spacecraft made of quartz and graphite.  We stress that the results of this paper can easily be modified for other material setups as necessary.  

We model the Starchip as a thin tube of height $H$, width $W$, and length $L$. {An optimal shape of the spacecraft perhaps is needle-like of $H=W\ll L$, as shown in Figure \ref{fig:model}. {Note that three different shapes of the spacecraft, including face on, edge on, and a long thin rod, were suggested by \cite{2016arXiv160401356L}.} We assume the spacecraft is moving with the velocity parallel to the long axis, such that the cross-section area of the spacecraft is $A=WH=H^{2}$. We perform our calculations assuming a relativistic speed for the spacecraft in the wide range $v=0.05c-0.5c$. 
%In this paper, we provide estimates for a fiducial spacecraft of 1 gram-mass with $L=5\cm, H=0.3\cm$ at solid density of $\rho\sim 2.2\g\cm^{-3}$

\begin{figure}
\centering
\includegraphics[width=0.3\textwidth]{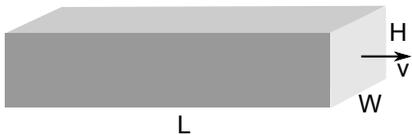}
\caption{Schematic of the needle-like spacecraft of height $H$, width $W$, and length $L$. The spacecraft surface area is $A=WH$.}
\label{fig:model}
\end{figure}

\section{Physical processes in relativistic spacecraft-gas interaction}\label{sec:physics}
We first discuss the important physical processes for a relativistic spacecraft in the ISM. The physics is analogous to that of relativistic dust presented in HLS15, but due to the much larger size of the spacecraft, some related effects are distinct from those of smaller relativistic grains.

\subsection{Bombardment by interstellar atoms}
Interstellar gas consists {mostly of hydrogen and helium with traces of heavy elements}. While H and He constitute most of the gas mass, massive atoms are more potent in causing damage to the spacecraft walls. We note that, since the electrons of the interstellar atom are rapidly stripped off upon entering the spacecraft surface, subsequent interactions of the atom with the target is essentially determined by its nucleus. In this paper, nucleus and ion are used interchangeably because the damage by atomic electrons is subdominant.

The interaction between a rapidly moving heavy ion and the surface can be divided into several phases. First, upon penetrating the surface, a fast heavy ion triggers numerous electronic excitations, mostly ionizations, which later produce energetic secondary electrons and holes. Before releasing secondary electrons, excited atoms with dense electronic excitations relax to low energy levels, accompanied by {\it lattice relaxation}, which involves the transfer of energy from highly excited atoms to nearby atoms. 

Next, these hot secondary electrons produce Auger electrons and quickly transfer their energy to lattice atoms in a narrow cylinder along the ion path, which transiently increases the temperature of the cylinder, establishing a {\it heating phase}. At this stage, a phase transition from solid to liquid can occur in the cylinder if the acquired temperature is above the melting point. As we discuss later, this leads to permanent defects if the cooling is sufficiently fast (i.e., liquid is quenched-in). 
Then, material in the cylinder cools down by transferring their energy to nearby atoms, leading to the {\it cooling phase}. Finally, excited atoms reach some equilibrium temperature of the lattice through {\it heat conduction}. The aforementioned processes occur on a short timescale of $10^{-13}-10^{-10}$ s at the microscopic level (see \citealt{2009JPCM...21U4205I}).
 
One important parameter characterizing energy transfer from a fast ion to the target is the rate of ion energy loss per unit length, also called stopping power, $dE/dx$. This is computed by summing over all possible ionizations and electronic excitations that the ion induces to target atoms (\citealt{1963ARNPS..13....1F}; \citealt{1999JAP....85.1249Z}; HLS15). 

\begin{figure*}
\includegraphics[width=0.45\textwidth]{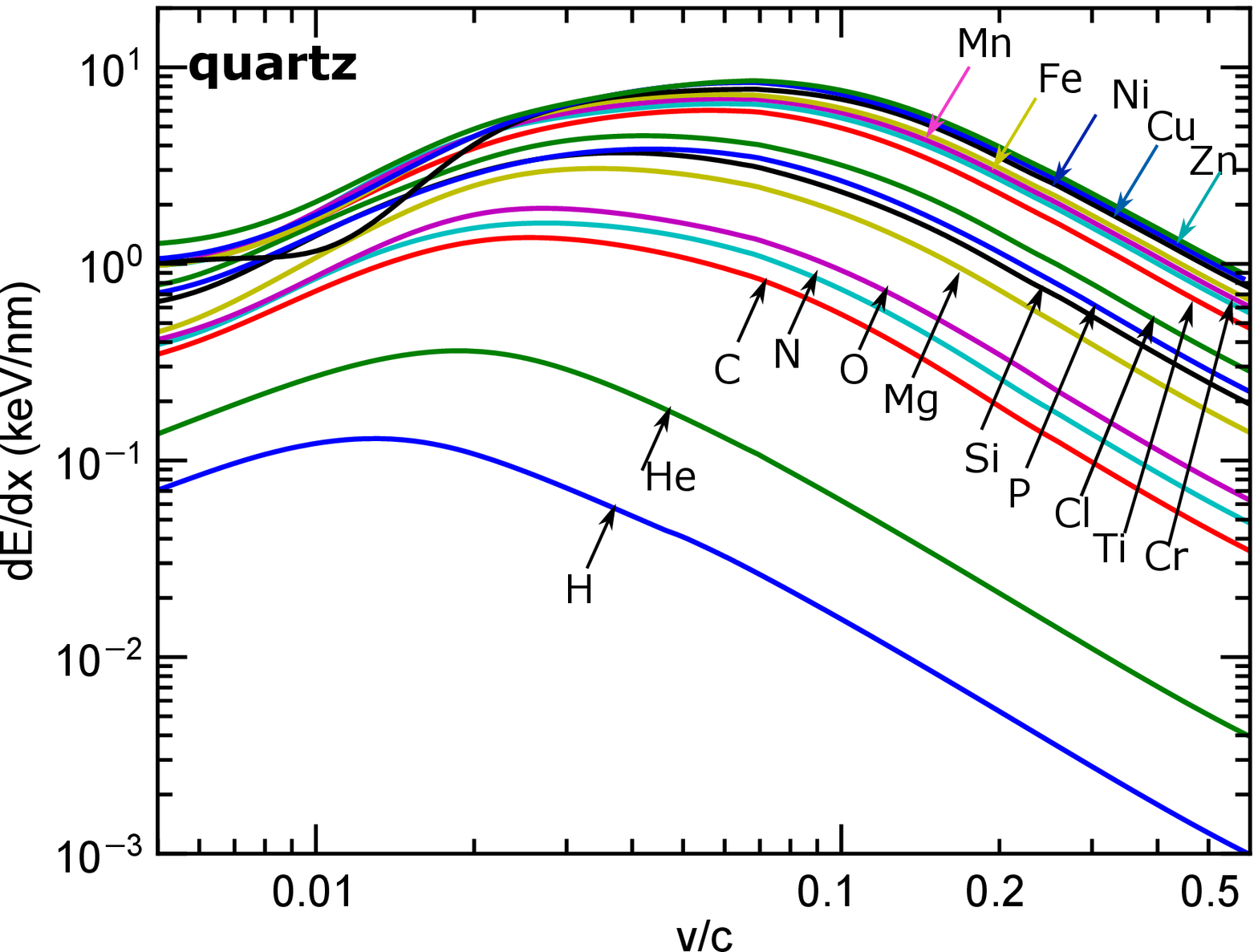}
\includegraphics[width=0.45\textwidth]{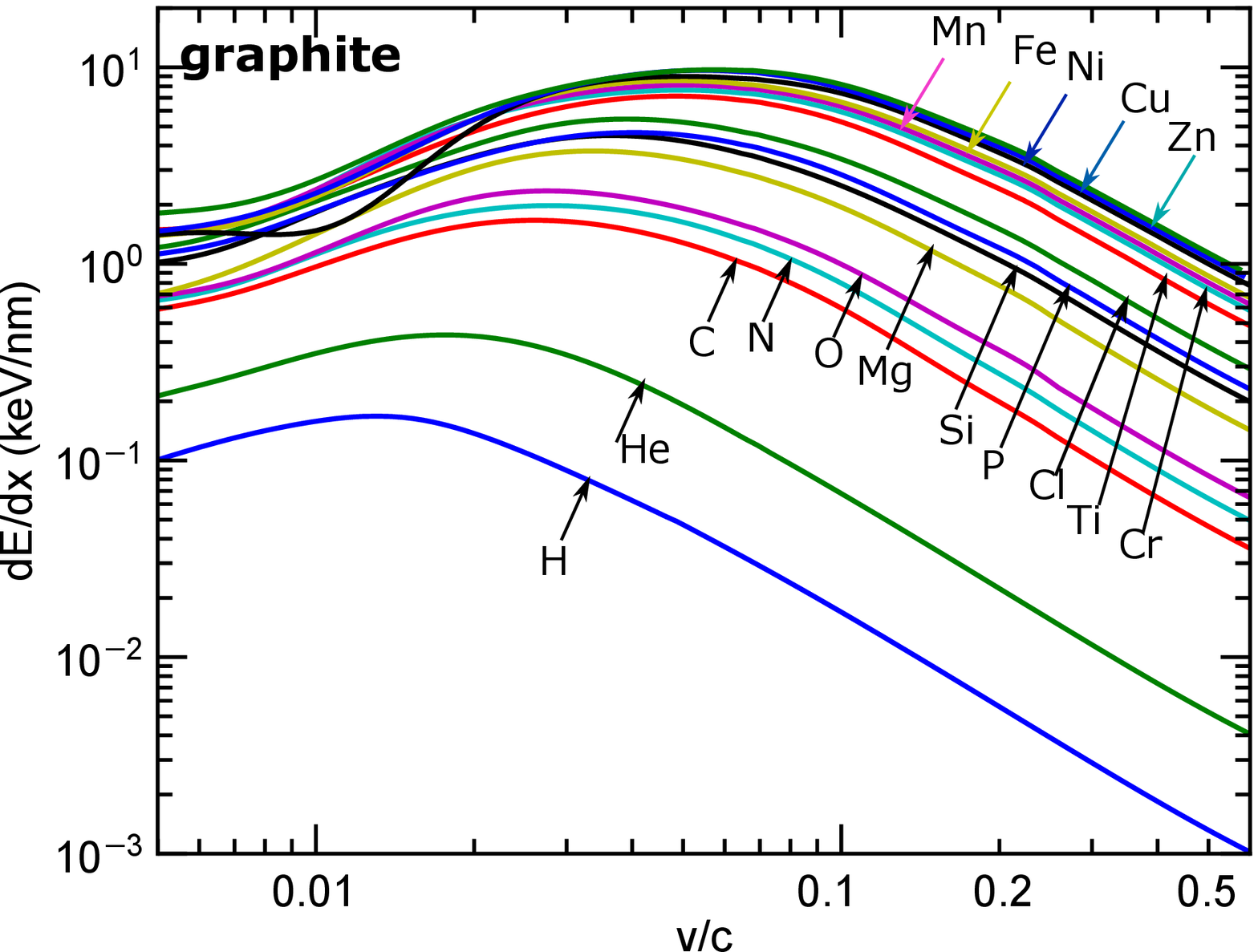}
\caption{Stopping power $dE/dx$ as a function of the ion speed, $v$, based on the SRIM software for interstellar gas elements with gas-phase abundance relative to hydrogen above $10^{-6}$. Both quartz (left panel) and graphite (right panel) materials are considered.}
\label{fig:dEdx}
\end{figure*}

To calculate stopping power $dE/dx$ for different interstellar atoms, we use the publicly available SRIM code (\citealt{2010NIMPB.268.1818Z}). SRIM allows us to compute $dE/dx$ from nuclear and electronic interactions for various ions and materials. Figure \ref{fig:dEdx} shows $dE/dx$ as a function of the ion speed for the 16 most abundant elements in the ISM, computed for quartz and graphite materials. 
The stopping power is maximal around some speed (e.g., $v\sim 0.015c$ for H), corresponding to the maximum cross-section of electronic interactions between ions and target atoms. When $v$ increases beyond the peak speed, $dE/dx$ falls rapidly because the cross-section of electronic interactions declines as $1/v^{2}$ (see e.g., HLS15). 
The value of $dE/dx$ for graphite is slightly larger than for quartz because the chosen graphite has higher atomic number density (see Table \ref{tab:material}).

{From Figure \ref{fig:dEdx} it follows that, at $v\sim 0.1c$, a light atom (He) deposits an excitation energy per target atom, 
$E_{\rm exc}=ldE/dx \le 6\times 10^{8}n^{-1/3} \sim 13$ eV where $n$ is the number density (see Table \ref{tab:material}) and $l=n^{-1/3}$ is the mean distance between two lattice atoms. Similarly, a heavy atom (Fe) deposits $E_{\rm exc}\sim 10^{3}$ eV. The energy provided by light atoms is sufficient to ionize a couple of electrons in the outer electronic shells { per target atom} following each collision with the spacecraft. Although these secondary electrons can transfer energy to lattice atoms, the low number of secondary electrons is insufficient to induce any modification in the structure material of spacecrafts. Heavy ions are therefore expected to produce major damage to the spacecraft due to much larger excitation energy.

\subsection{Formation of Damage Track}
The formation of permanent damage tracks of a few nanometer width in a solid by fast heavy ions was reported many years ago (\citealt{1959PMag....4..970S}; \citealt{1965JAP....36.3645F}). Since then, track formation has been extensively studied for various materials, including insulators \citealt{2004NIMPB.216....1T}, semiconductors \cite{1994PhRvB..4912457M}, graphite \citep{2001PhRvB..64r4115L}, and metals (\citealt{1994NIMPB..90..330D}). 

The physics of track formation is complex (see e.g., \citealt{2009JPCM...21U4205I} for a review). The basic idea is that, to form permanent damage to a solid, the lattice structure must be modified during the heating stage or lattice relaxation stage, provided that the cooling is sufficiently rapid so that the modified structure is quenched-in. 

Two leading models were proposed to explain track formation, namely the {\it thermal spike} model (\citealt{Seitz:1949fe}) and the {\it displacement spike} model \citep{1965JAP....36.3645F}. In the thermal spike model, track formation is thought to occur during the heating stage. In the displacement spike model, Coulomb repulsion between transiently ionized atoms in the hot cylinder directly converts electrostatic energy to atomic motion, resulting in the displacement of atoms away from the track core \citep{1982NIMPR.198..103J}. The thermal spike model can explain track formation in insulators \citep{2000NIMPB.166..903T} as well as in metals \citep{1994JPCM....6.6733W}. 

%For energetic heavy ions, all atoms in the hot cylinder track of length $l$ can be excited, and it is convenient to introduce an average excitation energy per atom as
%\bea
%E_{\rm exc} = \frac{ldE/dx}{n_{s}l\pi r_{\rm cyl}^{2}}=\frac{dE/dx}{n_{s}\pi r_{\rm cyl}^{2}},\label{eq:Eexc}
%\ena
%where $n_{s}$ is the atomic number density of the spacecraft material, and $r_{cyl}$ is the radius of the cylinder. 

Experimental studies show that the damage track is formed when the stopping power $dE/dx$ is larger than some threshold value $S_{\rm th}$. Such a threshold $S_{\rm th}$ varies with materials \citep{2009JPCM...21U4205I}. For quartz (SiO$_2$, an insulator), the threshold stopping power is $S_{\rm th}=1.5{\rm keV/nm}$ (\citealt{1994PhRvB..4912457M}). 
For graphite, track formation was observed for $S_{\rm th}\sim 5.1 {\rm keV/nm}$ \citep{2001PhRvB..64r4115L}. For material with high thermal conductivity (e.g., Cu, diamond), track formation is not expected at any stopping power. 

In Table \ref{tab:material} we present two well-known surface materials and their properties, including the mass density ($\rho$), atomic number density ($n$), threshold stopping power $S_{\rm th}$, binding energy $U_{0}$, and melting temperature $T_{m}$.

\begin{table}
\caption{Physical properties of spacecraft materials considered in this paper. }\label{tab:material}
\begin{tabular}{llllll} \hline\hline\\
{\it Material} & {$\mathcal{S}_{\rm th}$(keV/nm)} & {$\rho$($g\cm^{-3})$} & {$n(\cm^{-3})$} & {$U_{0}$(eV)}  & {$T_{m}$(K)}\\[1mm]
\hline\\
SiO$_2$ &1.5 & 2.32& 6.98E+22 & 6.4 &  1800\\[1mm]
Graphite & $5.1$ & 2.25 & 1.12E+23 & 4.0  &4000\\[1mm]
%BeCu & NA & 8.27 & 7.96e22 &  1358\\[1mm]
\hline\hline\\
\end{tabular}
\end{table}

%\subsection{Electronic sputtering}
An important process accompanied track formation by heavy ions is the sputtering of atoms from the spacecraft surface. HLS15 found that electronic sputtering is dominated by Fe, but it is rather inefficient at relativistic speeds $v\sim c$. For a spacecraft traveling at $v\sim 0.2$c, electronic sputtering is expected to be more efficient because of the higher stopping power (see Fig. \ref{fig:dEdx}). However, due to the low abundance of gas-phase Fe, sputtering has a minor effect for the damage of the spacecraft surface.

\subsection{Heating of spacecraft by gas collisions and radiation}
Light interstellar atoms essentially transfer most of their energy to target atoms, raising the temperature of the spacecraft. Heavy elements produce damage tracks as well as provide a source of heating. Interstellar photons and cosmic microwave background radiation can also heat the spacecraft. 

As we demonstrate in the next section, gas atoms at relativistic speeds are fully stopped over a distance of a few millimeter, much smaller than the centimeter-length of spacecrafts, transferring their entire energy to the surface layer. Thus, we have a situation in which the surface is supplied with a constant heat flux from collisions. This heat is then transferred inward through heat conduction and can raise the temperature of electronic devices, unless it is radiated away.  As we will show, some temperature difference between the front and back sides of the spacecraft might be a useful source of power for thermopower batteries on board the spacecraft.

\subsection{Charging of spacecraft}

For a cm-size spacecraft, incident electrons and ions will be stopped by the spacecraft and will not increase the overall charge. Although secondary electrons are created during this bombardment process, { they usually do not have enough energy to travel from the created place to the spacecraft surface to become free electrons}. Indeed, the range of the secondary electrons is rather short of $R_{e}\sim 120$~\AA~ for $E_{e}\sim 1$keV (see Equation \ref{eq:Rel}). As a result, collisional charging is negligible.

The photoelectric effect is still considerable because, for ultraviolet (UV) photons of energy below $100$ eV, the photon attenuation length (absorption length) is about 10\AA~ (\citealt{2001ApJS..134..263W}). Therefore, a fraction of photoelectrons will escape the surface. However, the acquired charge is insufficient to induce a major effect on the spacecraft. This is in contrast to the case of relativistic grains moving at $v\sim c$ where the UV radiation is boosted to X-ray, resulting in efficient dust destruction via Coulomb explosions or ion field emission (HLS15). 

\section{Effects of interactions with interstellar gas}\label{sec:ISMgas}
We now quantify the effects of spacecraft-relativistic gas interactions which were outlined in Section \ref{sec:physics}.

\subsection{Penetration length of interstellar atoms}
We seek to calculate the average length $R$ that an energetic atom can penetrate inside the spacecraft. The average length that a projectile of initial energy $E_{0}$ penetrates into solid before completely stopped is defined as
\bea
R(E_{0})=\int_{E_{0}}^{0}\left(\frac{dE}{dx}\right)^{-1}dE,\label{eq:RE}
\ena
where $dE/dx$ is the stopping power of the atom within the material. 

In calculating $R(E_{0})$, we take $dE/dx$ computed with the SRIM code, as shown in Figure \ref{fig:dEdx}. Figure \ref{fig:RE} (left) shows the derived penetration length as a function of atomic mass, $M$, for the different values of spacecraft speeds. Atoms with higher speeds can penetrate deeper into the solid, as expected. For a typical speed of $v=0.2c$, heavy ions are stopped within $R<1$ mm, while light ones (i.e., H, He) are stopped at larger depths. 

\begin{figure*}
\centering
\includegraphics[width=0.4\textwidth]{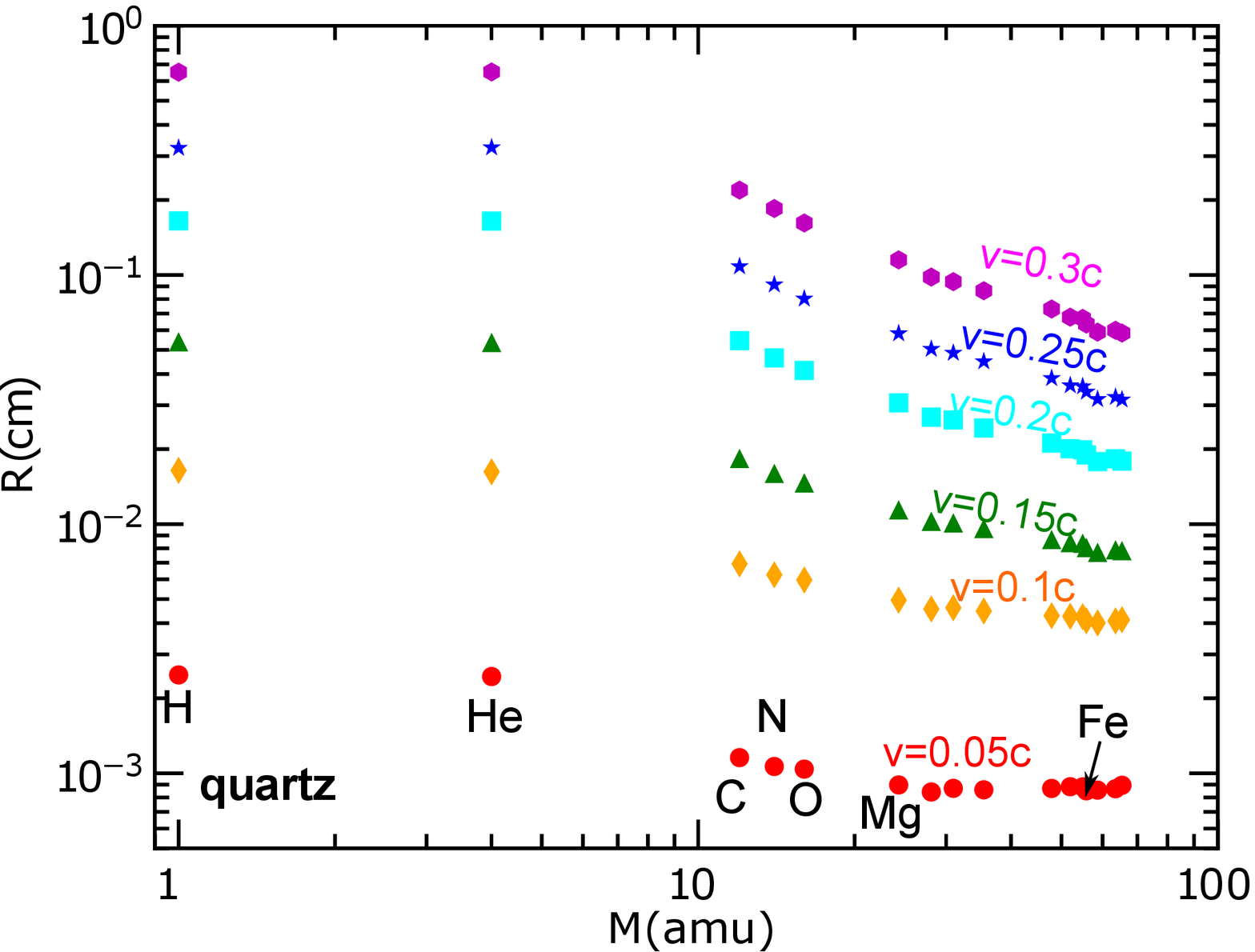}
\includegraphics[width=0.4\textwidth]{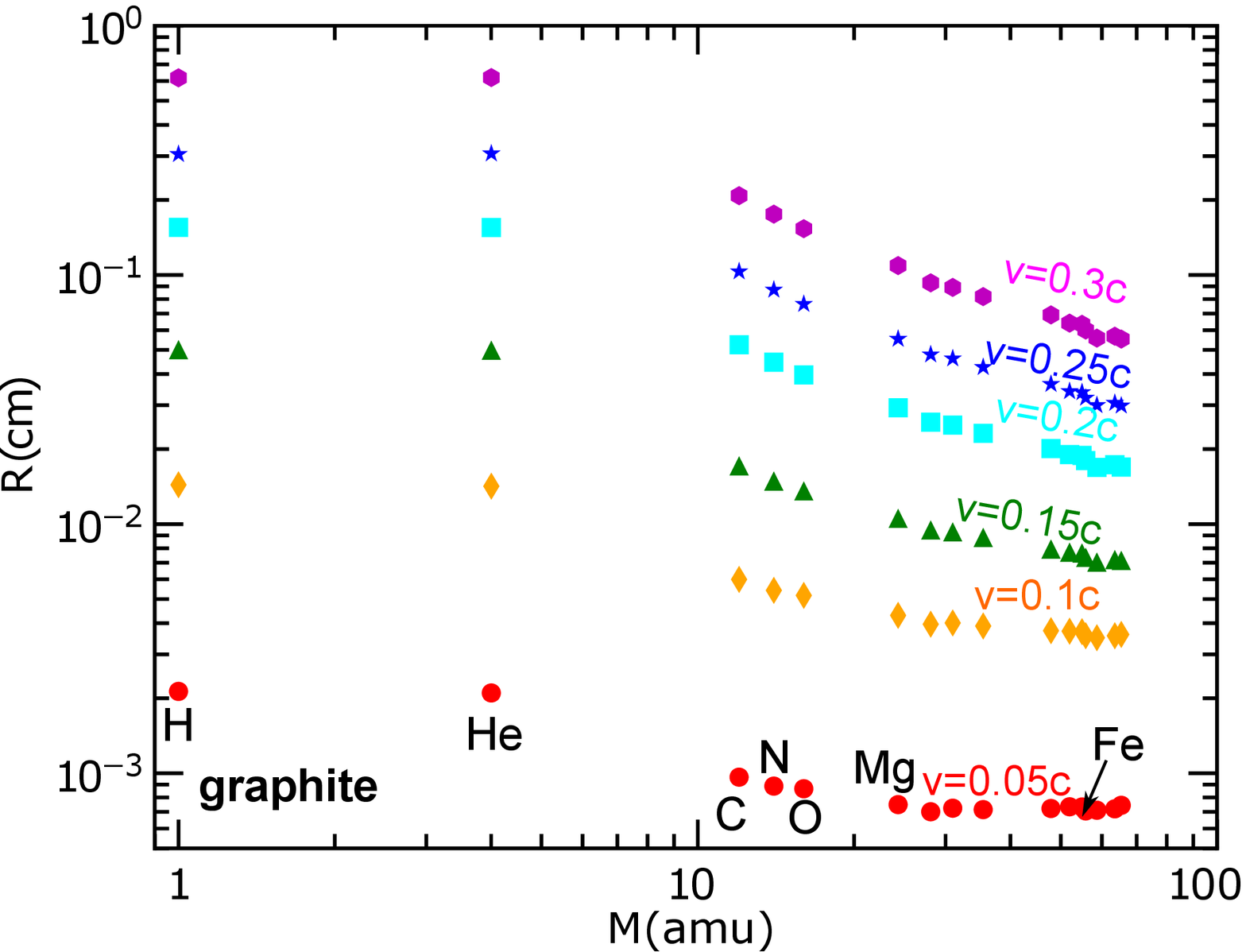}
\caption{Penetration length of the 16 most abundant elements in the ISM for several values of the spacecraft speed $v=0.05-0.3c$. Quartz and graphite material are considered in the left and right panels, respectively.}
\label{fig:RE}
\end{figure*}

\subsection{Track Radius}
The radius of a damage track produced by fast heavy ions has been measured for quartz (\citealt{1994PhRvB..4912457M}) and graphite \citep{2001PhRvB..64r4115L}. Analytical models were suggested to relate the track radius with the ion stopping power $dE/dx$, including the bond-breaking model \citep{1994NIMPB..94..424T} and thermal spike model \cite{1997NIMPB.122..530S}.

Following \cite{1997NIMPB.122..530S}, the radius of the ion track in quartz can be described by
\bea
r_{\rm tr}^{2}&=& a_{0}^{2}{\rm ln}\left(\frac{dE/dx}{{S}_{\rm th}}\right) {~\rm for~} {S}_{\rm th}<dE/dx<2.7{S}_{\rm th},\label{eq:rtr}\\
r_{\rm tr}^{2}&=& a_{0}^{2}\frac{dE/dx}{2.7{S}_{\rm th}} {~\rm for~} {dE/dx}>2.7{S}_{\rm th},\label{eq:rtr_high}\\
r_{\rm tr}&=& 0 {~\rm for~} dE/dx<{S}_{\rm th},
\ena
where $a_{0}$ is a model parameter, and ${S}_{\rm th}$ is the threshold power listed in Table \ref{tab:material}. 

\begin{figure*}
\centering
\includegraphics[width=0.4\textwidth]{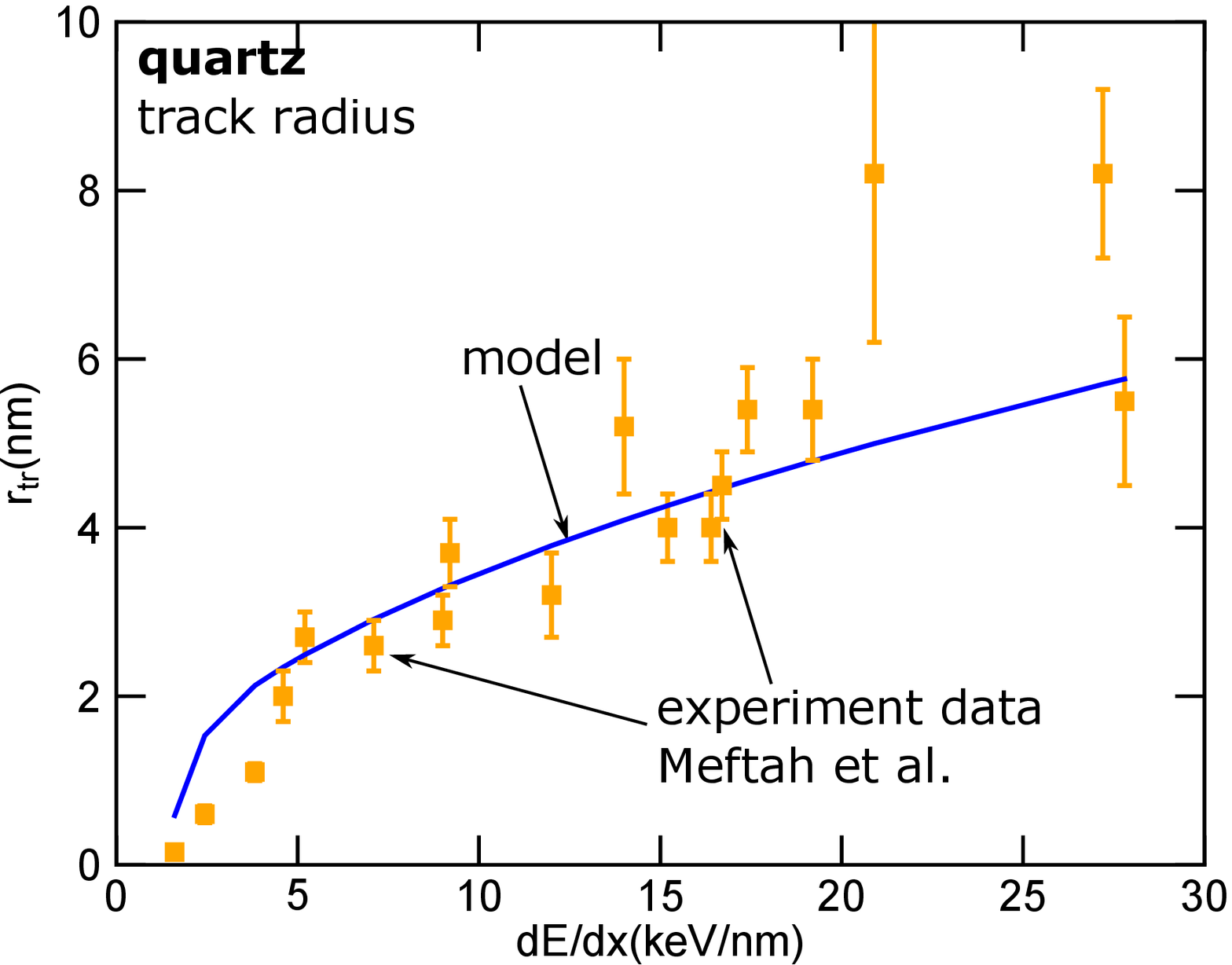}
\includegraphics[width=0.4\textwidth]{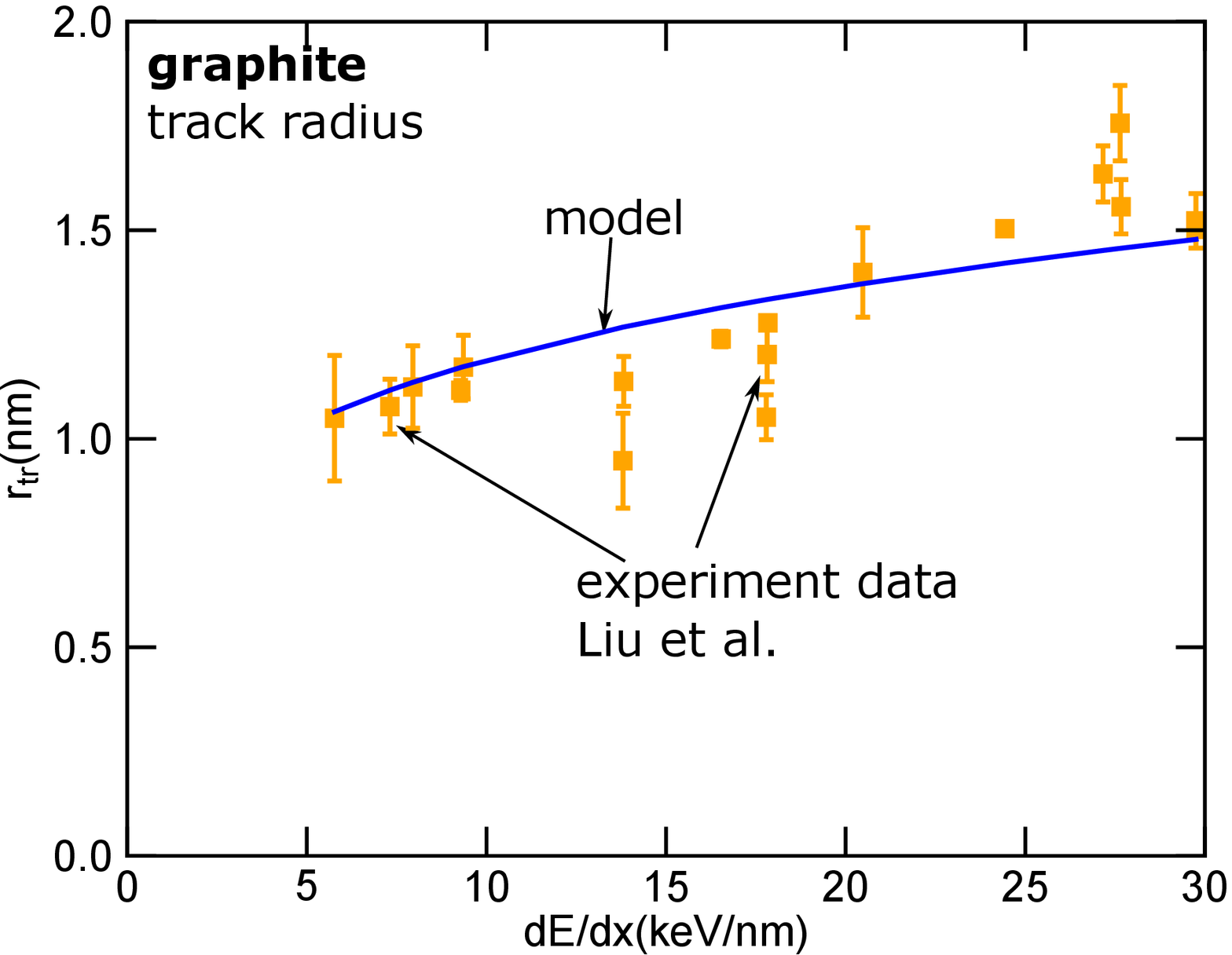}
\caption{Analytic fits to experimental data for SiO$_2$ (left panel, data from \citealt{1994PhRvB..4912457M}) and graphite (right panel, data from \citealt{2001PhRvB..64r4115L}). The data for quartz are fitted using Equations (\ref{eq:rtr}) and (\ref{eq:rtr_high}), while the data for graphite are fitted by a power law function $r_{\rm tr}\propto (dE/dx)^{\alpha}$ with $\alpha=0.2$.}
\label{fig:Rtr_exp}
\end{figure*}

Figure \ref{fig:Rtr_exp} presents the track radius as a function of $dE/dx$ computed with Equations (\ref{eq:rtr}) and (\ref{eq:rtr_high}) compared to the experimental data from \cite{1994PhRvB..4912457M}. The good fit is obtained for $a_{0}=2.2$ nm. In addition, the data for graphite can be fitted with a power law $r_{\rm tr}=a_{0}(dE/dx/{S}_{\rm th})^{\alpha}$. The fit is good for high $dE/dx$, but slightly overestimates the data for $dE/dx<10 {\rm keV/nm}$. For the same $dE/dx$, the track radius of graphite is smaller than of quartz because the latter is well conducting material that transfers heat from the track core faster.

To compute the track radius induced by various heavy ions in the ISM, we use Equation (\ref{eq:rtr}) with the best-fit parameters and interpolate for the ion stopping power $dE/dx$. Figure \ref{fig:Rtr_cal} (upper) shows the values of $r_{\rm tr}$ computed for the different atoms at several ion speeds from $v=0.05$c to $v=0.4c$.
For quartz, the track radius decreases significantly with increasing $v$ (upper panel). This stems from the fact that $dE/dx$ decreases rapidly with increasing $v$ (see Figure \ref{fig:dEdx}). For $v=0.05c$, track formation exists for ions with atomic mass $M\ge 16$ (oxygen and heavier atoms). At a higher speed of $v=0.1c$, the ion that produces a track should have a mass $M\ge 25$, and only ions with $M>55$ can produce track at $v=0.4c$. For graphite (lower panel), the track radius depends slightly on the speed for $v=0.05c-0.15c$, and no track is produced for $v\ge 0.2c$.

\begin{figure}
\includegraphics[width=0.45\textwidth]{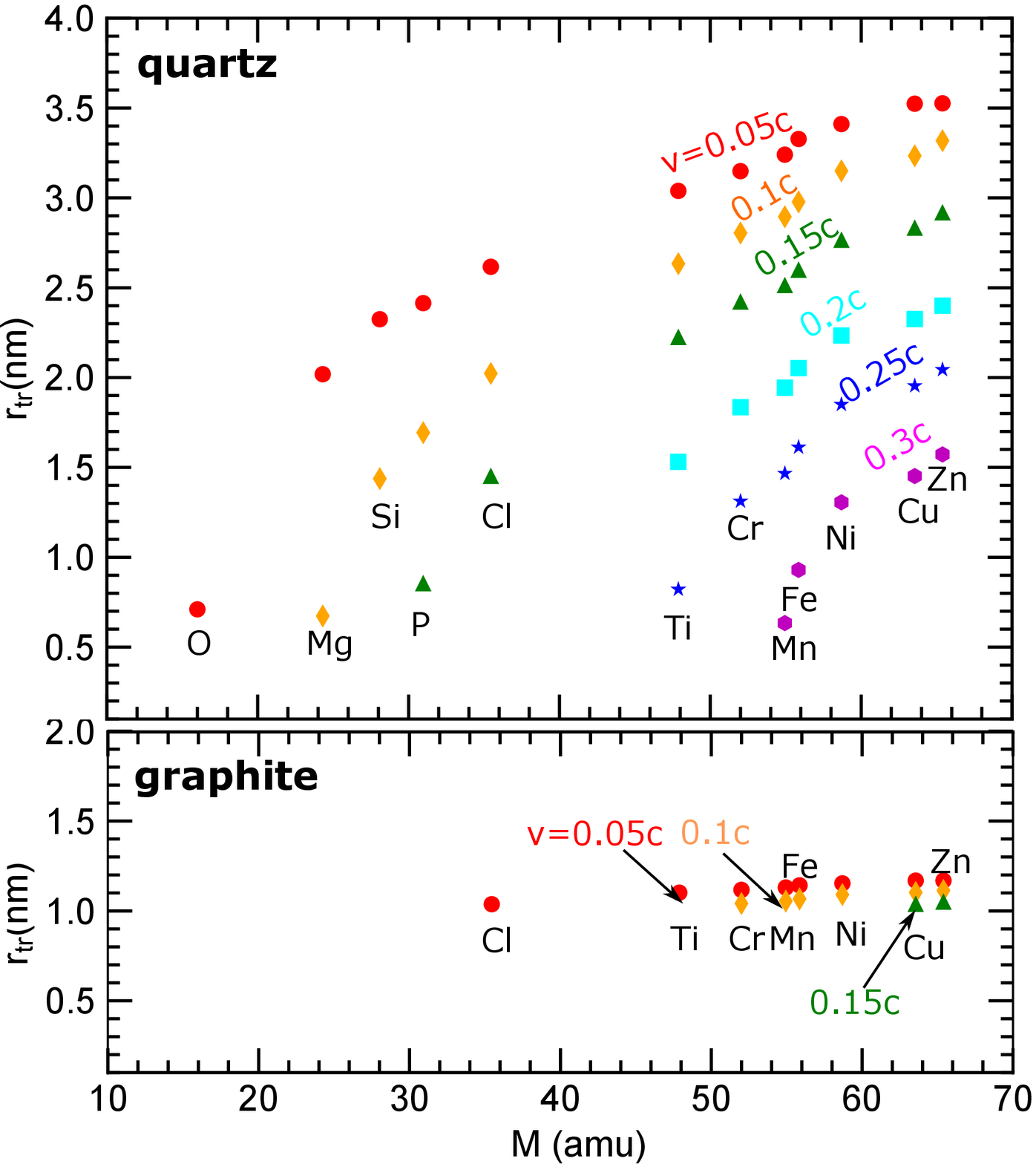}
\caption{Radius of damage track computed for the various ions bombarding a surface made of quartz (upper panel) and graphite (lower panel). Higher speeds result in lower track radius. Only very heavy ions (e.g. Fe) can produce tracks in graphite at $v\le 0.15c$.}
\label{fig:Rtr_cal}
\end{figure}

\subsection{Surface Damage}
Let $x_{i}$ be the gas-phase abundance of element $i$ relative to hydrogen, such that the density of element $i$ is $n_{i}=x_{i}n_{\H}$ where $n_{\H}$ is the proton number density in the gas. The collisional rate of the spacecraft with gas atoms $i$ is $n_{i}vA$ where $A$ is the geometrical cross-section of the spacecraft.
Each heavy atom $i$ produces a damage track of radius $r_{tr,i}$ and area $\pi r_{tr,i}^{2}$. Therefore, the total surface area of the spacecraft damaged by all interstellar atoms after a time interval $dt$ is calculated as the following:
\bea
dS =\sum_{i} \pi r_{tr,i}^{2}\times x_{i}n_{\H}vA \times dt=\sum_{i} \pi r_{tr,i}^{2} x_{i}AdN,~\label{eq:dSdl}
\ena
where $dN=n_{\H}vdt$ is the column density of gas swept by the spacecraft after $dt$. 

To obtain the total damaged surface area $S$ of the spacecraft after traversing the gas column $N_{\H}$, we can integrate Equation (\ref{eq:dSdl}) from $N=0$ to $N=N_{\H}$. However, there is possible overlaps between different tracks created at the different epochs. Thus, Equation (\ref{eq:dSdl}) must be multiplied with the probability that the incoming atoms do not fall in to the already damaged area, which is equal to $(A-S)/A$ (see \citealt{Gibbons:1972uw}).
The final version of Equation (\ref{eq:dSdl}) is therefore,
\bea
dS =\sum_{i} \pi r_{tr,i}^{2} x_{i}A\left(1-\frac{S}{A}\right)dN.\label{eq:dSdl_cor}
\ena

The fraction of the surface area damaged by gas collisions after traversing a gas column of $N_{\H}$ is calculated as:
\bea
f_{S}=\frac{\int_{0}^{S} dS}{A}
=1-\exp\left(-\sum_{i} \pi r_{tr,i}^{2}x_{i}N_{\H}\right),\label{eq:fsurf}~~~
\ena
where $r_{tr,i}$ is the function of the speed (see Figure \ref{fig:Rtr_cal}). { For the assumed geometry of the spacecraft (see Figure \ref{fig:starchip}), $f_{S}$ is independent of the spacecraft surface area $A$}.

The gas-phase abundance of some elements for the line of sight toward $\alpha$ Centauri was measured by \cite{1996ApJ...463..254L} who estimated $\log(x_{\rm Fe})\equiv \log(N_{\rm Fe}/N_{\H})=-5.05$ to $-5.65$, and $\log(N_{\rm Mg}/N_{\H})=-4.78$ to $-5.38$ for Fe and Mg, respectively. The gas-phase abundance of other elements is taken from \cite{2009ApJ...700.1299J}.

Figure \ref{fig:fsurf_quartz} (left panel) shows the value of $f_{S}$ as a function of $N_{\H}$ for different elements moving at $v=0.05c$, where the shaded line shows the range of measured gas column density toward $\alpha$ Centauri. The surface is damaged after the spacecraft sweeps a gas column $N_{\H}\sim 3\times 10^{17}\cm^{-2}$. The major contribution to the damage is from heavy elements such as O and Fe. 

Figure \ref{fig:fsurf_quartz} (right) shows the total value $f_{S}$ obtained from Equation (\ref{eq:fsurf}) for different speeds $v=0.05-0.3c$. The fraction of the damaged surface decreases with increasing $v$ because of the decrease of the track radius with $v$ (see Figure \ref{fig:Rtr_cal}, upper panel). For $v=0.2c$, about $70\%$ of the surface is damaged when $N_{\H}\sim 2\times 10^{18}\cm^{-2}$.

\begin{figure*}
\centering
\includegraphics[width=0.4\textwidth]{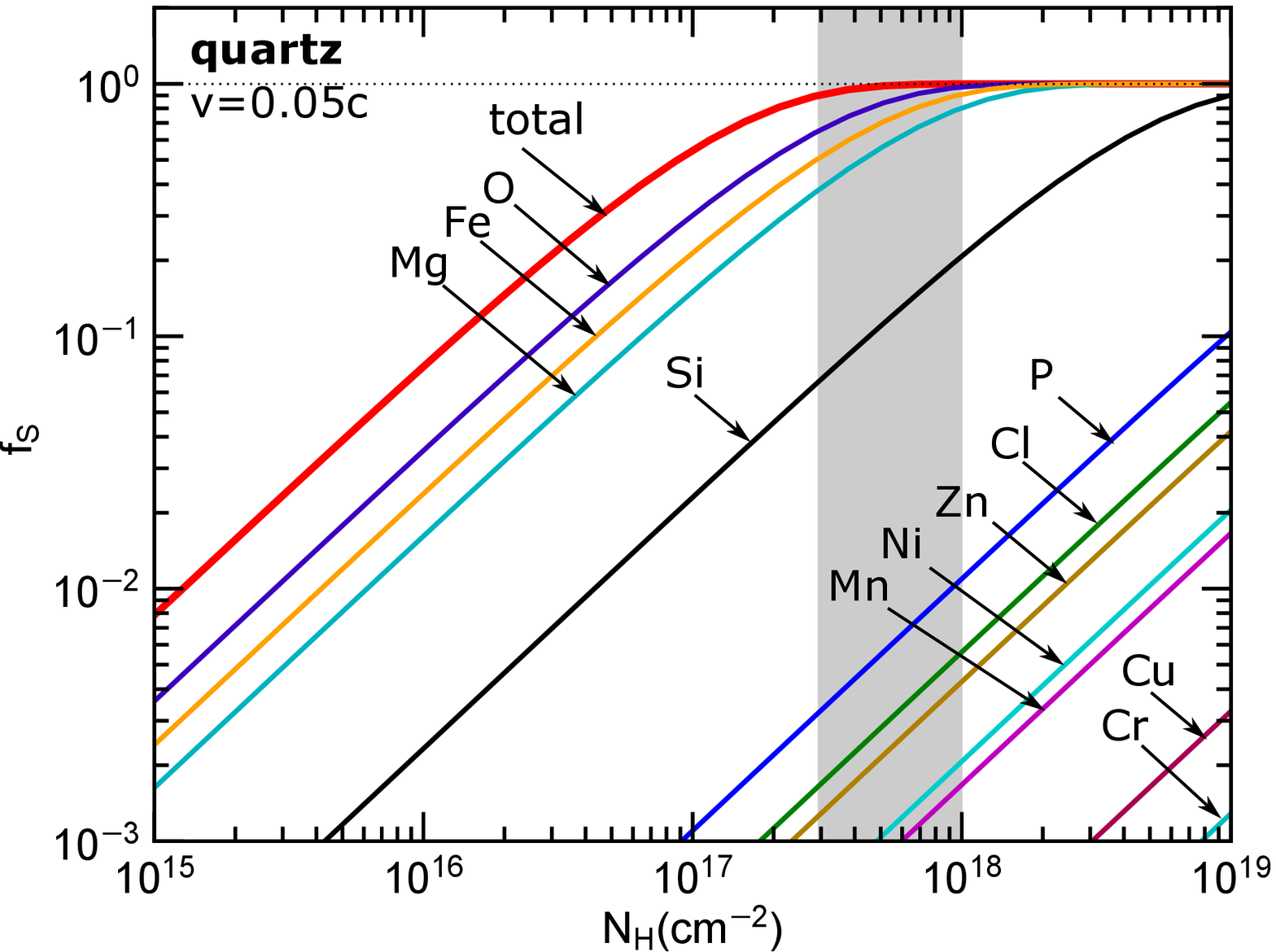}
\includegraphics[width=0.4\textwidth]{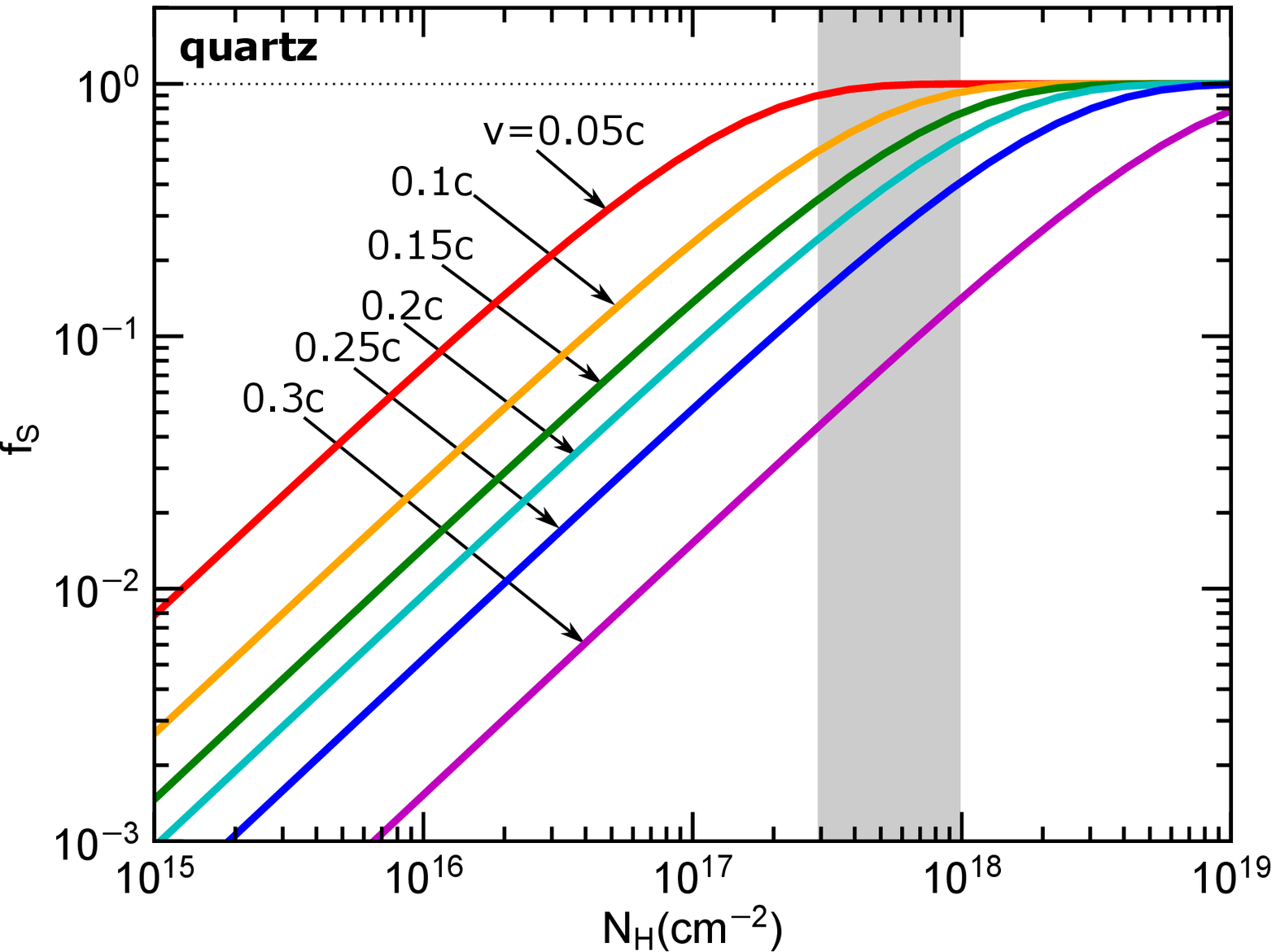}
\caption{Left panel: fraction of spacecraft surface damaged by gas bombardment as a function of the column density of gas swept by a spacecraft moving at $v=0.05c$. The contribution by each element is shown, with the dominant damage arising from O and Fe. The shaded area shows the measured gas column density toward $\alpha$ Centauri of $log N_{\rm H,obs}=17.80\pm 0.30$ (\citealt{1996ApJ...463..254L}). Right panel: total value $f_{S}$ for different speeds from $v=0.05c-0.3c$.}
\label{fig:fsurf_quartz}
\end{figure*}

Figure \ref{fig:fsurf_gra} shows the results for graphite. Here, the damage is small for $v=0.05c$, with $f_{S}\sim 0.3$ at $N_{\H}=10^{18}\cm^{-2}$. Above $v=0.1c$, the damage is negligible (see right panel).
 
\begin{figure*}
\centering
\includegraphics[width=0.4\textwidth]{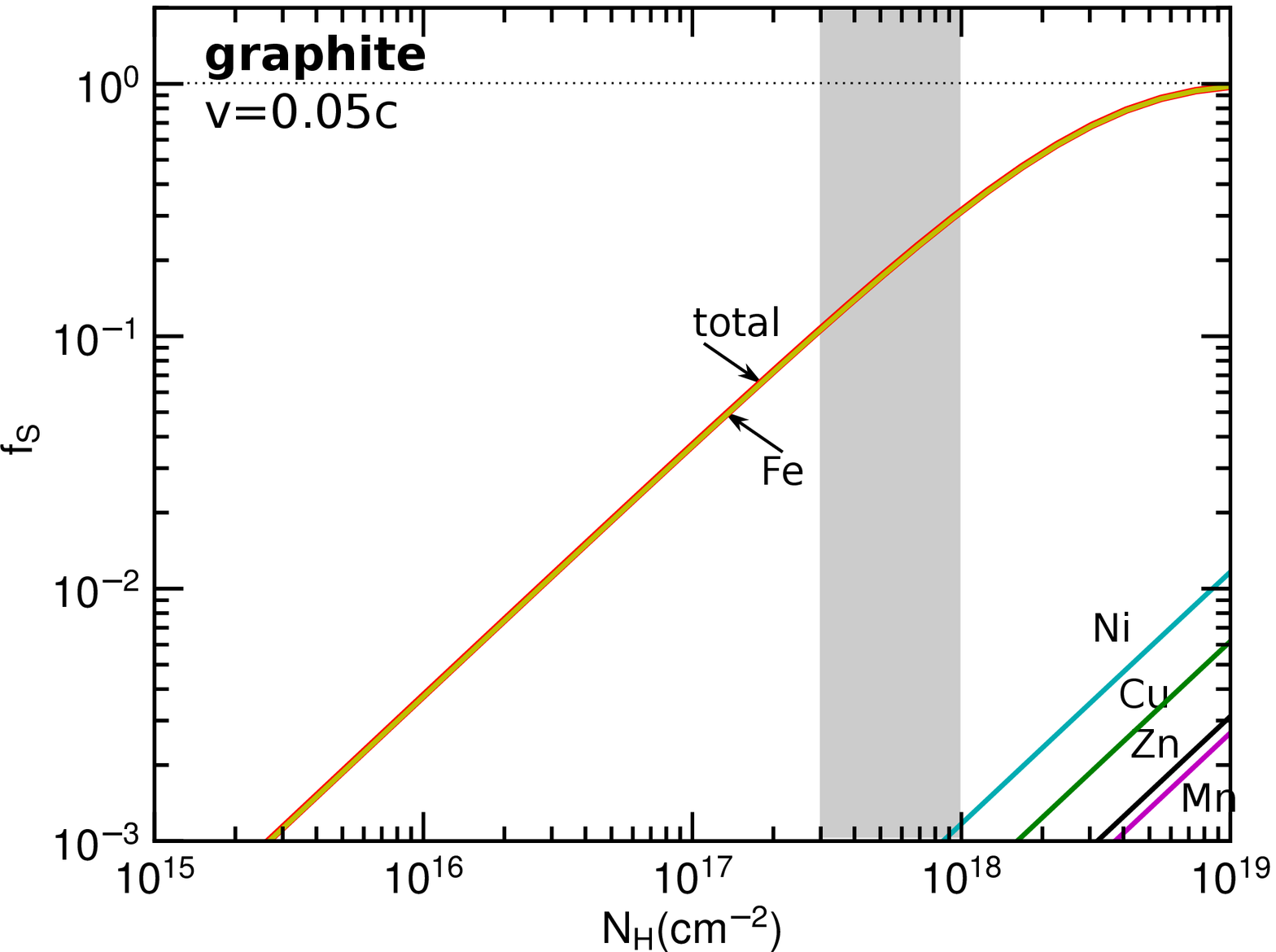}
\includegraphics[width=0.4\textwidth]{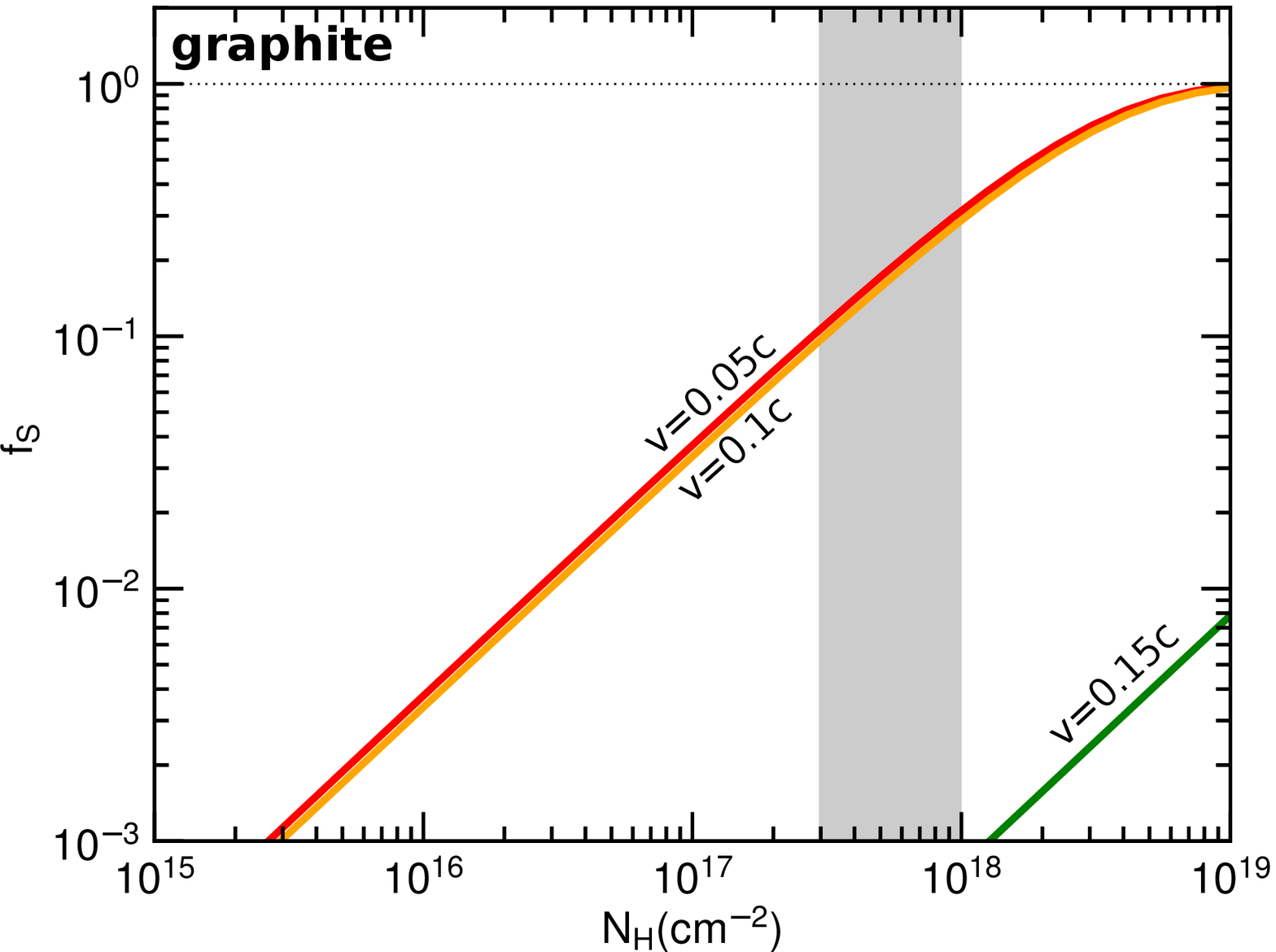}
\caption{Same as Figure \ref{fig:fsurf_quartz}, but for graphite material. Damage is essentially determined by Fe atoms.}
\label{fig:fsurf_gra}
\end{figure*}

\subsection{Volume filling factor of damage tracks}
The increase in the damaged volume of the spacecraft after $dt$ is given by
\bea
dV=\sum_{i}\pi r_{tr,i}^{2}R_{i}\times x_{i}n_{\H}vAdt \times \left(1-\frac{V}{LA} \right),\label{eq:dVdt}
\ena
where $R_i$ is the penetration length of element $i$, and the summation is taken over only ions that can produce tracks (i.e, $dE/dx\ge {S}_{\rm th}$). Here, the term $1-V/LA$ denotes the probability that the newly damaged volume will not coincide with the previously damaged volume. 

Equation (\ref{eq:dVdt}) can be rewritten as
\bea
\frac{dV}{1-V/LA} = \sum_{i}\pi r_{tr,i}^{2}R_{i}\times x_{i}A dN.
\ena

Finally, we calculate the volume filling factor of damage tracks by gas bombardment:
\bea
f_{V}=\frac{\int_{0}^{V} dV}{LA}
%=\sum_{i}\frac{\pi r_{tr,i}^{2}R_{i}x_{i}N_{\H}}{L},\\
=1-\exp\left(-\frac{\sum_{i}\pi r_{tr,i}^{2}R_{i}x_{i}N_{\H}}{L}\right).\label{eq:fV}~~
\ena

\begin{figure*}
\centering
\includegraphics[width=0.4\textwidth]{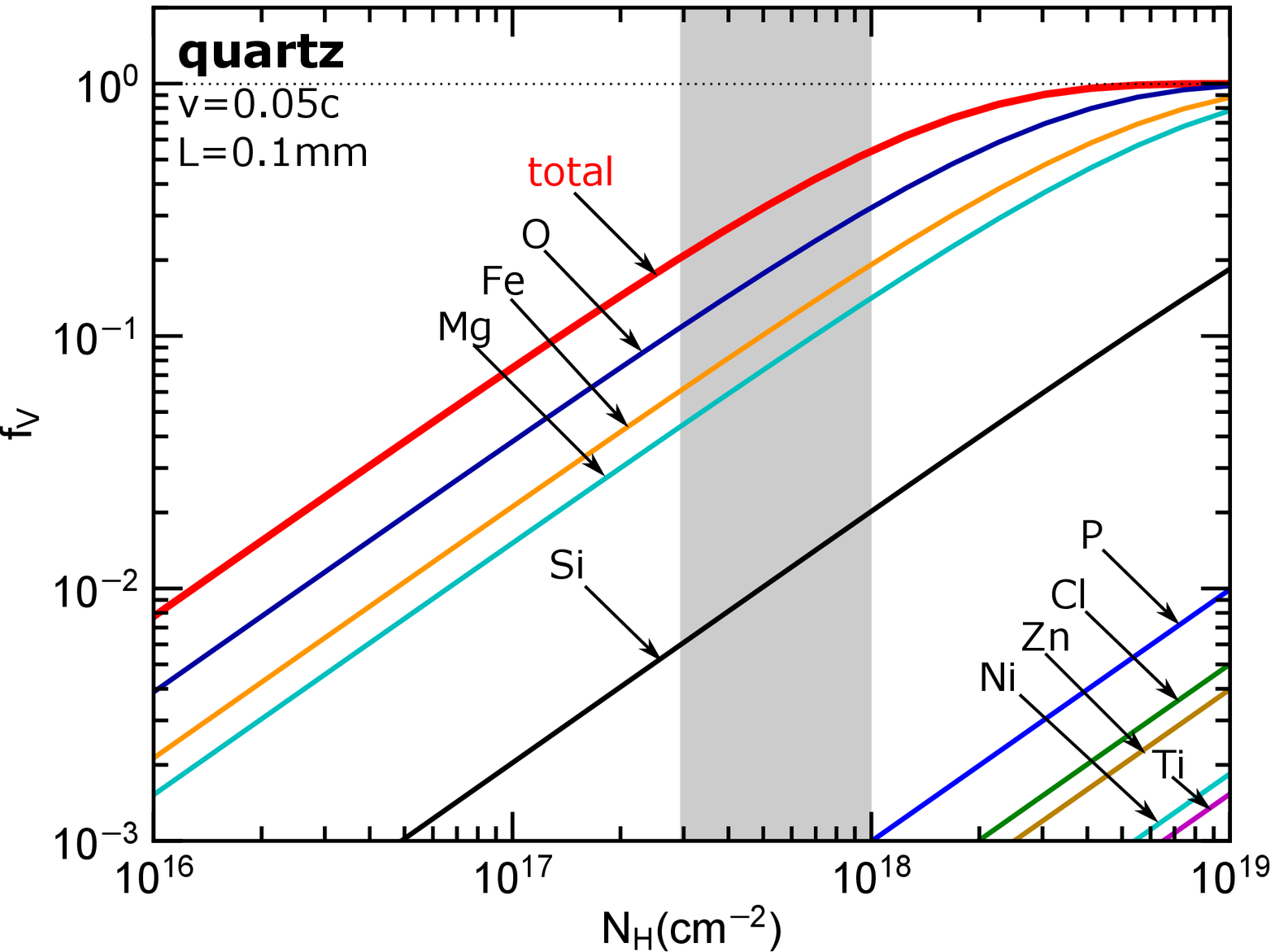}
\includegraphics[width=0.4\textwidth]{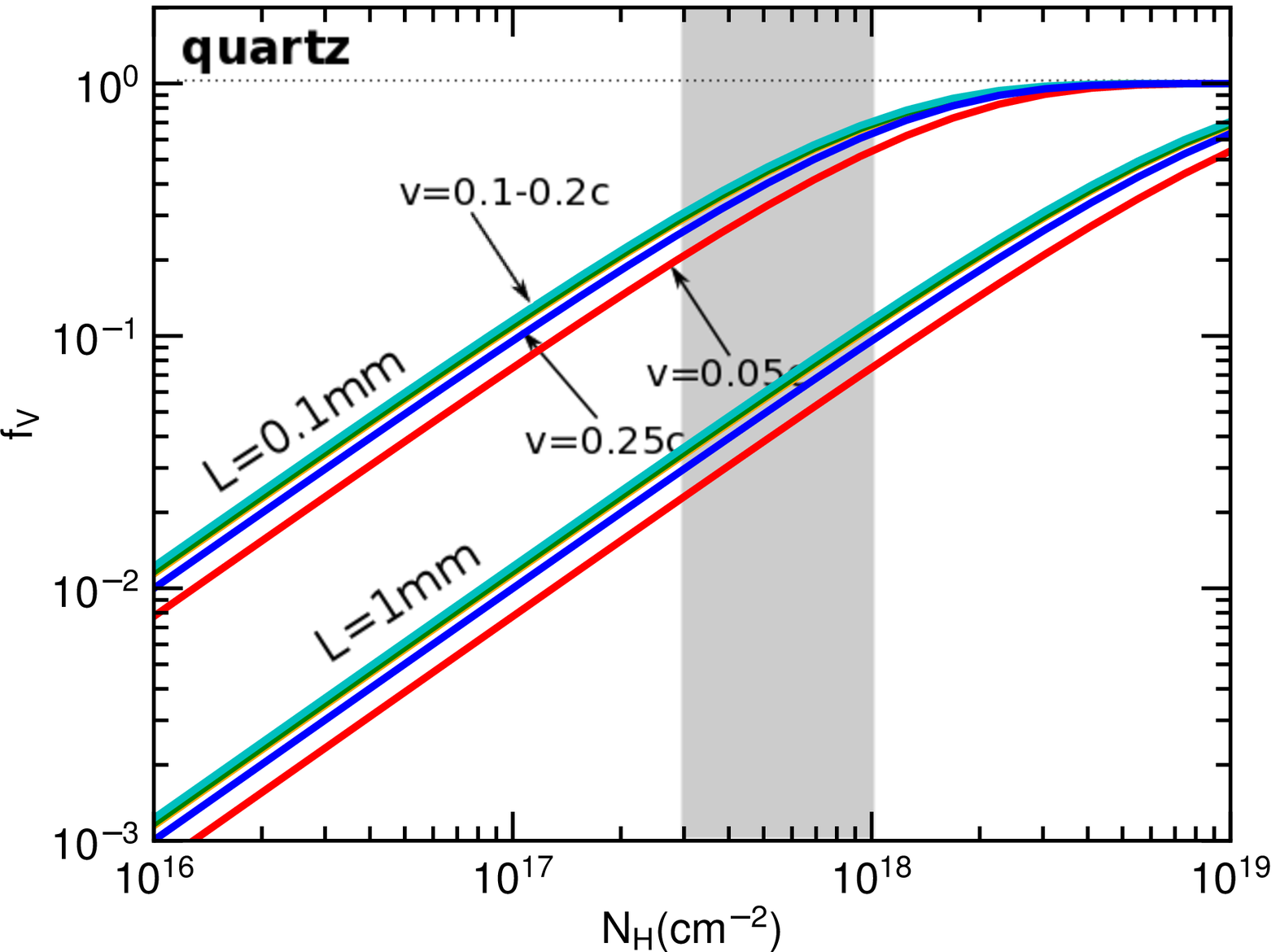}
\caption{Left panel: volume filling factor of damage tracks in a spacecraft of thickness $L=0.1$ mm by different ions. Right panel: total volume filling factor of damage tracks for the different speeds with $L=0.1$ mm and $L=1$mm. Quartz material is considered.}
\label{fig:fV}
\end{figure*}

The left panel of Figure \ref{fig:fV} shows $f_{V}$ due to different interstellar atoms at $v=0.05c$. The damage is dominated by O, Fe, and Mg, despite their low abundance. A surface layer of thickness $L=0.1$ mm can be damaged substantially after the spacecraft has swept a gas column of $N_{\H}\sim 10^{18}\cm^{-2}$. The right panel of Figure \ref{fig:fV} shows the results for different speeds $v=0.05-0.25c$ and the thickness of $L=0.1$ mm and 1 mm. Interestingly, the value of $f_{V}$ varies slowly with $v$, in contrast to the surface damage $f_{S}$ (see Figure \ref{fig:fsurf_quartz}). This is because the increasing penetration length can compensate for the decrease of the surface area at larger speeds $v$.

Figure \ref{fig:fV_gra} shows the similar results for graphite. In difference from quartz, graphite can be damaged to a very thin layer of $L\sim 0.01$ mm by the time the spacecraft reaches $\alpha$ Centauri.

\begin{figure}
\includegraphics[width=0.4\textwidth]{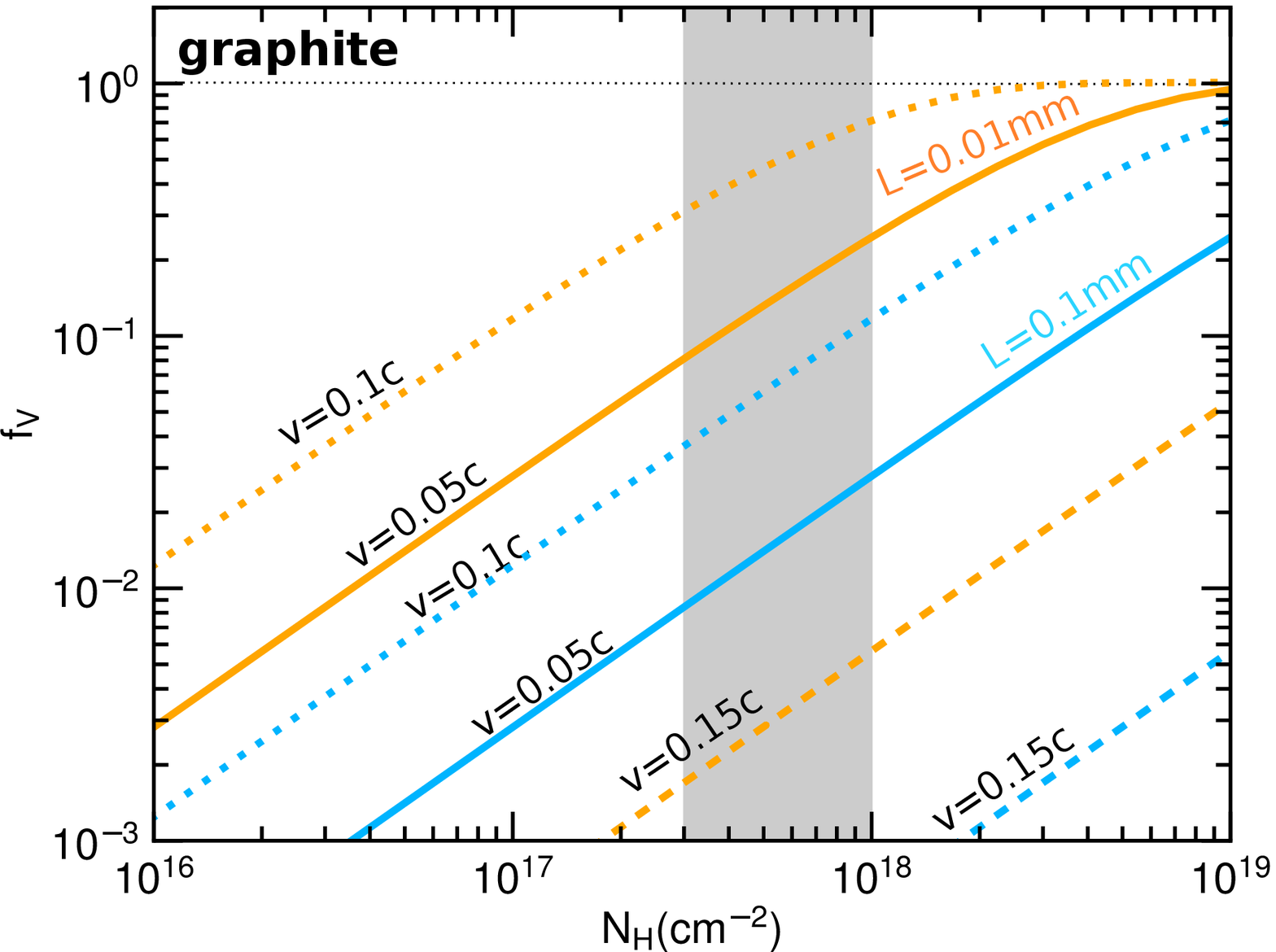}
\caption{Same as Figure \ref{fig:fV} but for graphite. Orange lines and bright blue lines show the results for $L=0.01$ mm and $L=0.1$ mm, respectively.}
\label{fig:fV_gra}
\end{figure}

\section{Interaction of a relativistic spacecraft with interstellar dust}\label{sec:ISMdust}

\subsection{Overview of possible consequences from dust bombardment}
In this section we consider the interaction of the relativistic spacecraft with interstellar dust grains. In the limit of relativistic speeds, dust bombardment to the target spacecraft can be considered as the simultaneous bombardment of a cluster of heavy atoms from the dust grain because the kinetic energy of atoms is much larger than the binding energy of the grain atoms. During the collision, grain atoms and target atoms are both ionized, producing energetic secondary electrons. 
Such hot electrons quickly loose their kinetic energy by transferring to target atoms in a cylinder along the grain track which raises the lattice temperature (see Section \ref{sec:physics}). As time goes on, the temperature declines while the hot cylinder spreads due to heat conduction (see Appendix \ref{apdx:heatcond}). This results in the damage of an extended area of the spacecraft surface. Naturally, the dust grain melts and evaporates gradually.

The energy loss per unit of pathlength of a dust grain of size $a$ to the target can be evaluated as
\bea
\frac{dE_{d}}{dx}=N_{d}\frac{dE_{1}}{dx}=\frac{4\pi a^{3}n_{d}}{3}\frac{dE_{1}}{dx},\label{eq:dEgr_dx}
\ena
where $N_{d}$ is the total number of atoms in the dust grain, $n_{d}$ is the atomic number density of the dust, and $dE_{1}/dx$ is the average energy loss per grain atom per pathlength in the spacecraft. To compute $dE_{1}/dx$, we adopt MgFeSiO$_4$ and pure C for interstellar silicate and graphite dust.

Let $\epsilon$ be the fraction of energy loss $dE_{d}/dx$ transferred to lattice atoms. The total energy transferred from the grain to target atoms in a cylinder of radius $R_{\rm cyl}$ and length $l$ is given by: 
\bea
\Delta E &=& \epsilon\frac{ldE_{d}}{dx}=\frac{4\pi a^{3}n_{d}}{3}\frac{\epsilon ldE_{1}}{dx},\nonumber\\
&\simeq & 4\times 10^{17}a_{-5}^{3}\left(\frac{n_{d}}{10^{23}\cm^{-3}}\right)\left(\frac{\epsilon ldE_{1}/dx}{10^{10}\rm eV}\right) \eV,~~~~\label{eq:N_dEdx}
\ena
where $a_{-5}=(a/10^{-5}\cm)$. {The exact value of $\epsilon$ is uncertain. Simulations show that about $\sim 60-80\%$ of $dE/dx$ went to electron kinetic energy (see \citealt{1994PhRvB..4912457M}). Since some fraction of the electron kinetic energy is spent in radiation, in the following, we take $\epsilon=0.5$ as a conservative value.}

To evaluate the damage induced by this huge energy $\Delta E$, we first need to calculate the instantaneous temperature of the heated cylinder. In the high temperature limit, the specific heat capacity from the Debye model reads $C_{V}= (3 N-6)k\sim 3Nk$ where $N=n_{s} \pi R_{\rm cyl}^{2}l$ is the total number of atoms in the cylinder and $n_{s}$ is the atomic number density of the spacecraft. 

As a result, the instantaneous temperature can be estimated as \footnote{Immediately after a collision, the temperature in the interaction cylinder tends to decrease with increasing distance from the cylinder core, and decrease with time as the cylinder spreads out due to heat conduction. The temperature is averaged over the rectangular cross-section of the affected cylinder (see Appendix \ref{apdx:heatcond}) for more details.}
\bea
T_{\rm cyl}&=&\frac{\Delta E}{C_{V}}=  \left(\frac{a}{R_{\rm cyl}}\right)^{3}\frac{4R_{\rm cyl}}{9}\frac{\epsilon dE_{1}/dx}{k},\nonumber\\
&\simeq & 5\times 10^{13}\epsilon\left(\frac{a}{R_{\rm cyl}}\right)^{3}\left(\frac{R_{\rm cyl}}{1\cm}\right)\left(\frac{dE_{1}/dx}{10^{10}\eV\cm^{-1}}\right) \K,~~~\label{eq:Td_coll}
\ena
where $n_{s}\sim n_{d}\approx 10^{23}\cm^{-3}$ is assumed for both the dust grain and spacecraft. We also assume that all ion energy loss is converted to heat, even though some small fraction of secondary electron energy is converted to Bremsstrahlung radiation.

When $T_{\rm cyl}$ exceeds the evaporation temperature $T_{\rm evap}=U_{0}/3k$, the overheated matter rapidly changes to vapor state, resulting in complete evaporation (see e.g., \citealt{1994ApJ...431..321T}). For $T_{\rm cyl}\sim T_{m}$, the heated matter is melted, changing from solid to liquid state.

Figure \ref{fig:starchip} presents a schematic illustration of the interaction between a micron grain with the spacecraft and a snapshot of the modification of the surface structure. Atoms in a limited volume are heated above the binding energy, such that they escape suddenly from the surface. The final outcome is an empty crater on the spacecraft surface. Material in a more extended cylinder is melted. We investigate these scenarios in the next subsections.

\begin{figure}
\includegraphics[width=0.45\textwidth]{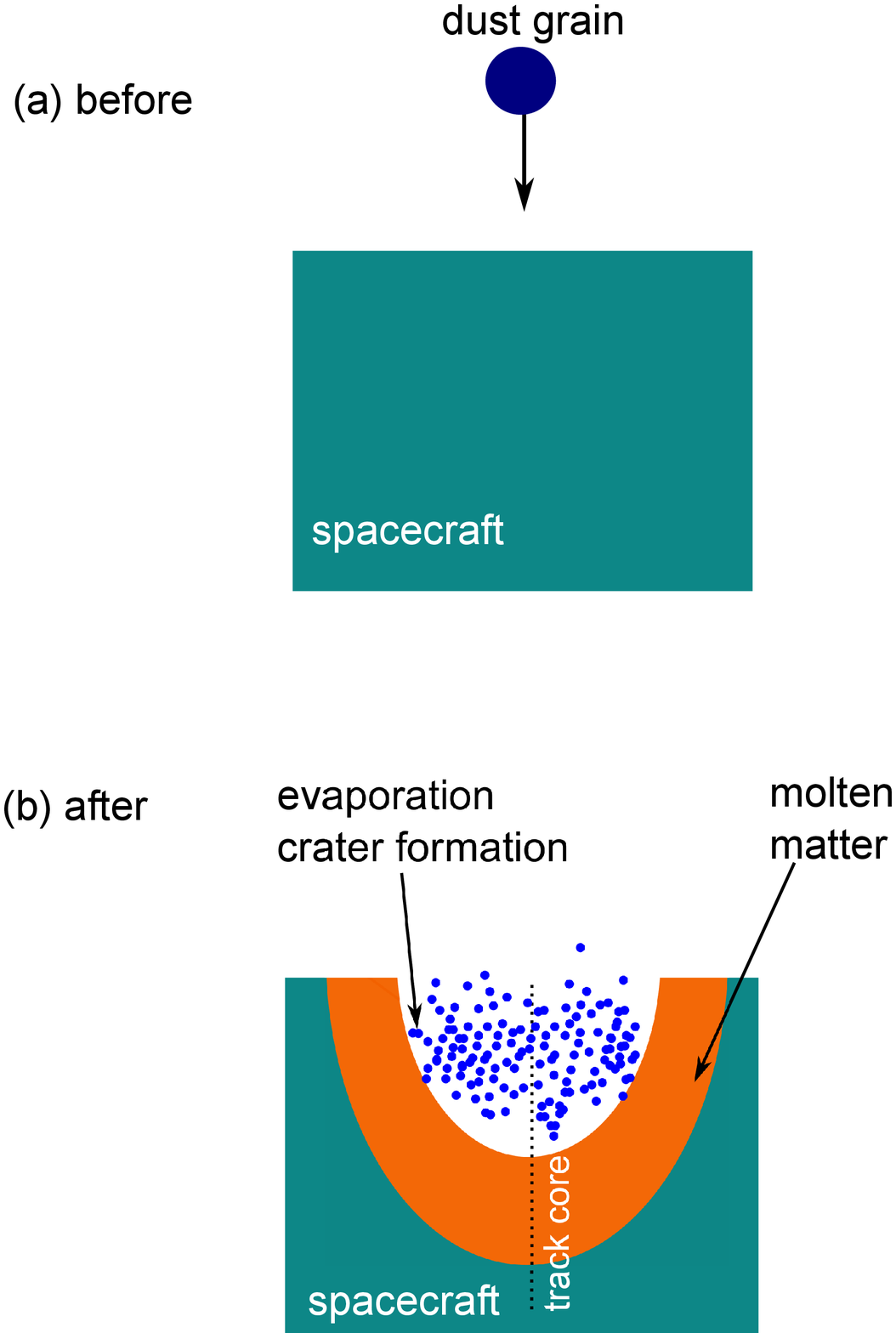}
\caption{Schematic illustration of the collision impact from a dust grain on the spacecraft surface considered in the spacecraft's rest frame.}
\label{fig:starchip}
\end{figure}

\subsection{Expected damage to the spacecraft by dust bombardment}
\subsubsection{Grain size distribution}
Since the effect of dust collision crucially depends on the size of dust grains, it is important to know the size distribution of interstellar grains. Current models of interstellar dust (e.g., \citealt{2001ApJ...548..296W}) show that most grain mass is concentrated below $\sim 0.25 \mum$, and very little dust is present above $0.3\mum$. {We note that the size distribution of interstellar dust toward Alpha Centauri is very uncertain, thus we will use the well-known model of \cite{2001ApJ...548..296W} to evaluate the expected damage.  As new data comes in, these results can be easily updated.} 

For our calculations, we assume that interstellar dust grains have the following size distribution:
\bea
\frac{dn_{j}}{n_{\H}da}&=&\frac{C_{j}}{a}\left(\frac{a}{a_{t,j}}\right)^{\alpha_{j}}F(a;\beta_{j},\alpha_{j})G(a;\beta_{j},\alpha_{j}),\label{eq:dnda}
\ena
where where $j=sil, gra$ for silicate and graphite compositions, $a_{t,j}$, $a_{c,j}$ are the model parameters, and $C_{j}$ is a constant determined by the total gas-to-dust mass ratio $R_{g/d}$  (see \citealt{2001ApJ...548..296W} for more detail). 

The coefficient functions $F$ and $G$ read:
\bea
F&=& 1+\beta a/a_{t} {~\rm for~} \beta>0,\\
F&=&(1-\beta a/a_{t})^{-1} {~\rm for~} \beta<0,
\ena
and
\bea
G(a;\beta_{j},\alpha_{j})&=&1 {~\rm for~} a<a_{t,j}\label{eq:dnda},\\
G(a;\beta_{j},\alpha_{j})&=&\exp\left(-[(a-a_{t,j})/a_{c,j}]^{3}\right) {~\rm for~} a>a_{t,j}.
\ena

For a standard model of the diffuse ISM with the total-to-selective extinction ratio $R_{V}=3.1$, we adopt the parameters from \cite{2001ApJ...548..296W}. The value of the gas-to-dust mass ratio is $R_{g/d}\sim 100$ for the ISM. This size distribution of both interstellar silicate and graphite drops exponentially for $a>0.25\mum$, so we adopt the upper cutoff of the size distribution $a_{\max}=1\mum$, and the lower cutoff $a_{\min}=0.001\mum$, unless explicitly stated otherwise.
  
\subsubsection{Formation of Craters due to Explosive Evaporation}
For sub-micron size dust grains, the heat by the grain collision mainly induces transient spot heating, resulting in sudden evaporation of a small volume (e.g., cylinder), and creates a hole or a crater on the spacecraft surface. 

The rate of collisions of the spacecraft with the dust component $j$ is given by
\bea
R_{\rm coll}(a)=n_{j}(a)v A,\label{eq:Rcoll}
\ena
where $n_{j}(a)=dn_{j}/da$, and the cross-section of the dust grain which is much smaller than $A$ has been disregarded. 

Let $R_{\rm cyl,evap}$ be the radius of the hot cylinder that is heated to $T_{\rm evap}$. To find $R_{\rm cyl,evap}$, we set $T_{\rm cyl}=T_{\rm evap}$ in Equation (\ref{eq:Td_coll}) and obtain
\bea
\pi R_{\rm cyl,evap}^{2}=\frac{\epsilon dE_{d}/dx}{n_{s}U_{0}}=\frac{4\pi a^{3}\epsilon dE_{1}/dx}{3U_{0}},\label{eq:Rcyl_evap}
\ena
where $n_{d}\sim n_{s}$ is the number density of dust. Plugging relevant numerical parameters, we obtain
\bea
R_{\rm cyl,evap}\simeq15 a_{-5}^{3/2}\left(\frac{\epsilon dE_{1}/dx}{1\rm keV nm^{-1}}\right)^{1/2}\left(\frac{6{\rm eV}}{U_{0}}\right)^{1/2}\mum.~~~~~
\ena

From Equation (\ref{eq:Rcoll}), the surface area of the spacecraft eroded by dust bombardment in a time interval $dt$ is (see Equation \ref{eq:dSdl}):
\bea
dS_{\rm evap}= \sum_{j}\int_{a_{\min}}^{a_{\max}}n_{j}(a)davA\times \pi R_{\rm cyl,evap}^{2}\times  dt,\label{eq:dSevap}
\ena
where $j=sil,gra$ for silicate and graphite compositions of the interstellar dust.

The fraction of the spacecraft surface area evaporated after traversing a column $N_{\H}$ is then equal to
\bea
\frac{f_{S,{\rm evap}}}{N_{\H}}&=&\sum_{j}\int_{a_{\min}}^{a_{\max}} \frac{R_{\coll}(a)}{A}\frac{4\pi a^{3}}{3}\frac{dE_{1}/dx}{U_{0}} \frac{dn_{j}}{da} da,\nonumber\\
&=&\sum_{j}\int_{a_{\min}}^{a_{\max}} \frac{4\pi a^{3}}{3}\frac{\epsilon dE_{1}/dx}{U_{0}} \frac{dn_{j}}{da} da,\label{eq:fsurf_evap}
\ena
where $j = sil, gra$. {We note that explosions result in the sudden loss of spacecraft material, so the effect of overlap between different collisions can produce deeper craters, such that the total area of craters can exceed the surface area $A$ of the spacecraft.}
%(Alex: not really so. We get deeper craters...)

\subsubsection{Melting and Modification of Material Structure}

Due to heat conduction, atoms in an extended cylinder with radius $r>R_{\rm cyl,evap}$ can be heated to the melting point $T_{m}$ (see Appendix \ref{apdx:heatcond}), so that matter within the cylinder is melted. Above the melting point, thermal sublimation \citep{1989ApJ...345..230G} can occur, resulting in the loss of the spacecraft mass. In addition to radiative cooling, thermal sublimation induces evaporative cooling, which reduces the sublimation rate. The subsequent evolution of the molten matter is essentially determined by thermal sublimation, thermal radiation, and heat conduction. 

For insulators (e.g., quartz), heat conduction is less efficient than sublimation and radiative cooling. For highly conducting material like graphite, heat conduction is important, suppressing the efficiency of sublimation. HLS15 found that thermal sublimation is inefficient because of rapid cooling by radiation and evaporation. As a result, the molten matter will cool down to below the melting point before another collision with interstellar dust {because the mean time between two successive collisions is much longer than the cooling timescales}. Therefore, the main consequence of melting is to induce the modification of the material structure, i.e., the newly-established structure will not be the same as the initial state. Here we assume that it is the protective surface of the spacecraft that is melted. Naturally, electronic devices heated above the melting point would lose their functionality.
%Alex: (??? not clear what is the result. depends on dust density: reply: time between two collisions is much longer than the cooling time)

The radius of the melting cylinder can be evaluated by Equation (\ref{eq:Rcyl_evap}) where $U_{0}$ is replaced by $3kT_{m}$:
\bea
\pi R_{\rm cyl,m}^{2}=\frac{\epsilon dE_{d}/dx}{3n_{s}kT_{m}}=\frac{4\pi a^{3}\epsilon dE_{1}/dx}{9kT_{m}}. \label{eq:Rcyl_melt}
\ena

%Melting by dust collision does not result in the erosion of the spacecraft surface. However, similar to energetic heavy ions that produce damage tracks, melting results in the modification of the structure of the spacecraft material and affect electronic components. Thus, it is important to also evaluate the damage via melting.

The fraction of melting surface area can be approximately estimated as in Equation (\ref{eq:fsurf_evap}):
\bea
f_{S,m}=1 - \exp\left( -N_{\H}\sum_{j}\int_{a_{\min}}^{a_{\max}} \frac{4\pi a^{3}}{9}\frac{\epsilon dE_{1}/dx}{kT_{m}} \frac{dn_{j}}{da} da\right),\label{eq:fsurf_melt}~~~~~
\ena
where the overlap between molten cylinders is accounted for. 

\begin{figure*}
\centering
\includegraphics[width=0.45\textwidth]{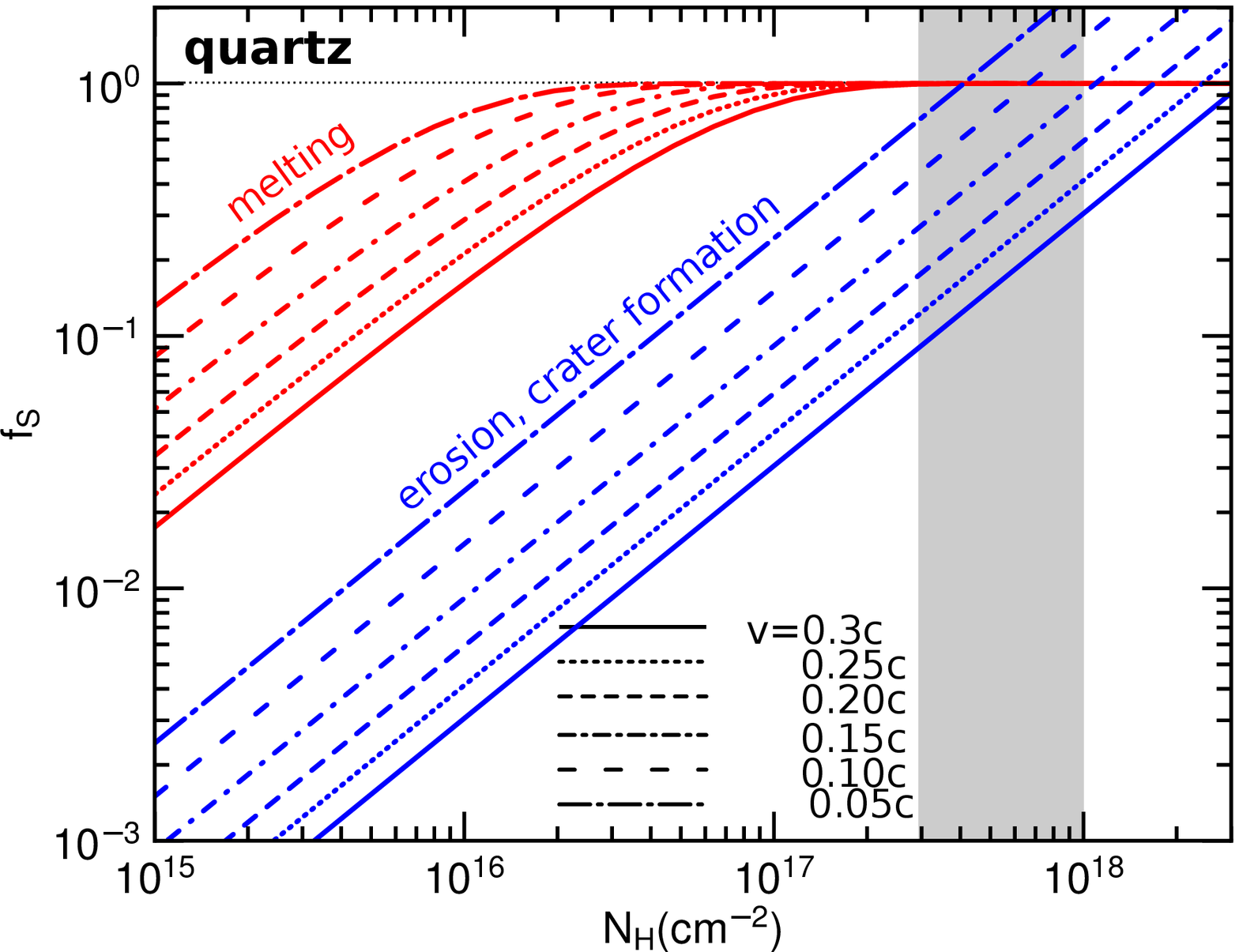}
\includegraphics[width=0.45\textwidth]{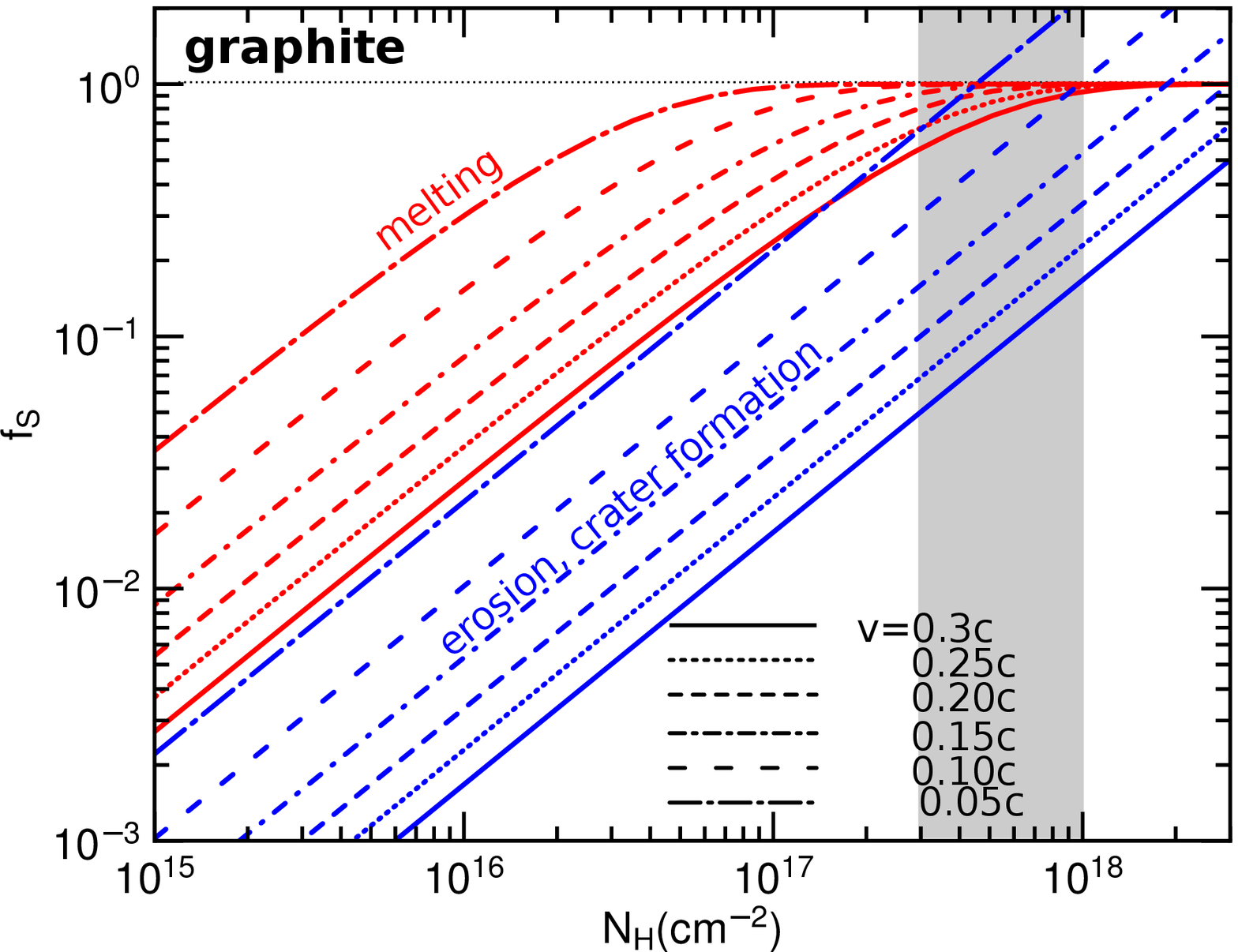}
\caption{Fraction of the spacecraft surface eroded (solid lines) and melted (dashed lines) due to dust bombardment, computed for quartz (left panel) and graphite (right panel) materials. The shaded areas show the measured gas column density toward $\alpha$ Centauri of ${\rm log} N_{\rm H,obs}=17.80\pm 0.30$ from \cite{1996ApJ...463..254L}.}
\label{fig:fsurf_dust}
\end{figure*}

Figure \ref{fig:fsurf_dust} shows the fraction of the spacecraft surface that is evaporated (red lines) and melted (blue lines) due to dust bombardment for quartz (left panel) and graphite (right panel). At $v=0.2c$, we find that about $20\%$ of the surface is eroded after a gas column $N_{\H}\sim 3\times 10^{17}\cm^{-2}$ for both quartz and graphite, which are well below the observed values toward $\alpha$ Centauri (see shaded region).  Melting is the most efficient for quartz surface, which melts the surface after $N_{\H}\le 10^{17}\cm^{-2}$ for the considered speeds. Melting for graphite is less efficient and requires $N_{\H}\ge 3\times 10^{17}\cm^{-2}$.

\subsubsection{Spacecraft volume eroded by dust bombardment}

Next we estimate the total volume of the craters formed by interstellar dust grains. A dust grain of size $a$ can heat all atoms in some volume $\delta V(a)$ to an average energy equal to the binding energy $U_{0}$,
\bea
n_{s}\delta V(a)U_{0}=\epsilon_{V} E_{d}=\frac{4\pi a^{3}n_{d}}{3} \frac{\epsilon_{V} mv^{2}}{2},\label{eq:dV_crater}
\ena
where $m$ is the average mass of grain atoms, and $\epsilon_{V}$ is the fraction of the grain kinetic energy $E_{d}$ converted to kinetic energy of secondary electrons which will go into lattice heating. We conservatively assume $\epsilon_{V}=0.5$.

The ratio of the total volume of craters to the spacecraft volume $LA$ is then obtained by integrating $\delta V(a)$ over the grain size distribution and collision rate:
\bea
\frac{f_{V,{\rm evap}}}{N_{\H}}&=&\sum_{j}\int_{a_{\min}}^{a_{\max}} \frac{R_{\coll}(a)}{LA}\frac{4\pi a^{3}}{3}\frac{\epsilon_{V} mv^{2}}{2U_{0}} \frac{dn_{j}}{da} da,\\
&=&\sum_{j}\int_{a_{\min}}^{a_{\max}}\frac{4\pi}{3L}\frac{\epsilon_{V} mv^{2}}{2U_{0}} \frac{a^{3}dn_{j}}{da} da,~~~ \label{eq:fvol_dust}
\ena
where $j=sil, gra$, and $n_{s}\sim n_{d}$ has been used. 

Similarly, we can evaluate the volume filling factor of molten material as follows:
\bea
f_{V,m}=1-\exp\left(- N_{\H}\sum_{j}\int_{a_{\min}}^{a_{\max}}\frac{4\pi}{3L}\frac{\epsilon_{V} mv^{2}}{6kT_{m}} \frac{a^{3}dn_{j}}{da} da\right),~~~\label{eq:fvol_dust_melt_cor}
\ena
where the overlap of molten cylinders is accounted for melting (cf., explosive evaporation).

\begin{figure*}
\centering
\includegraphics[width=0.45\textwidth]{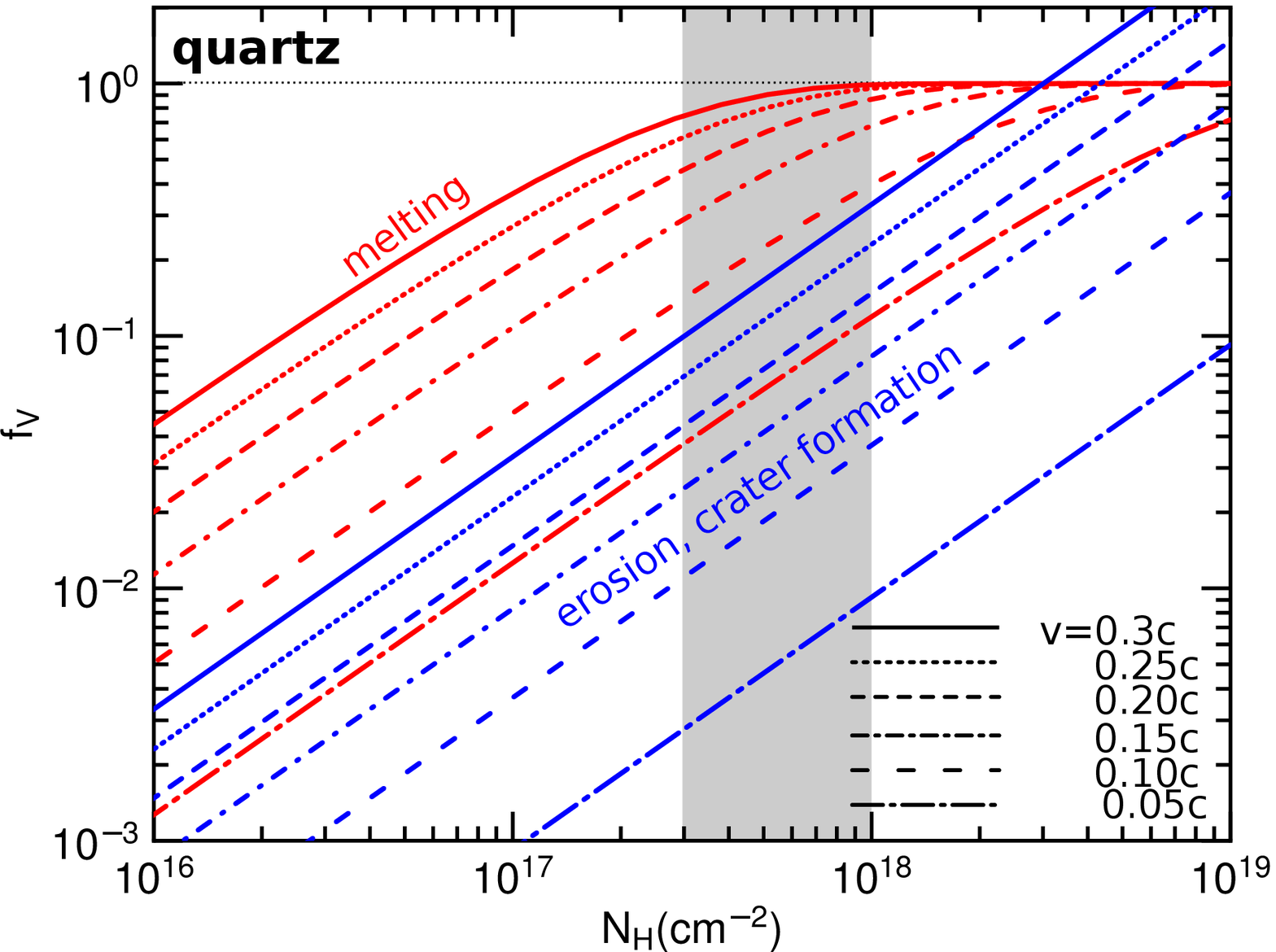}
\includegraphics[width=0.45\textwidth]{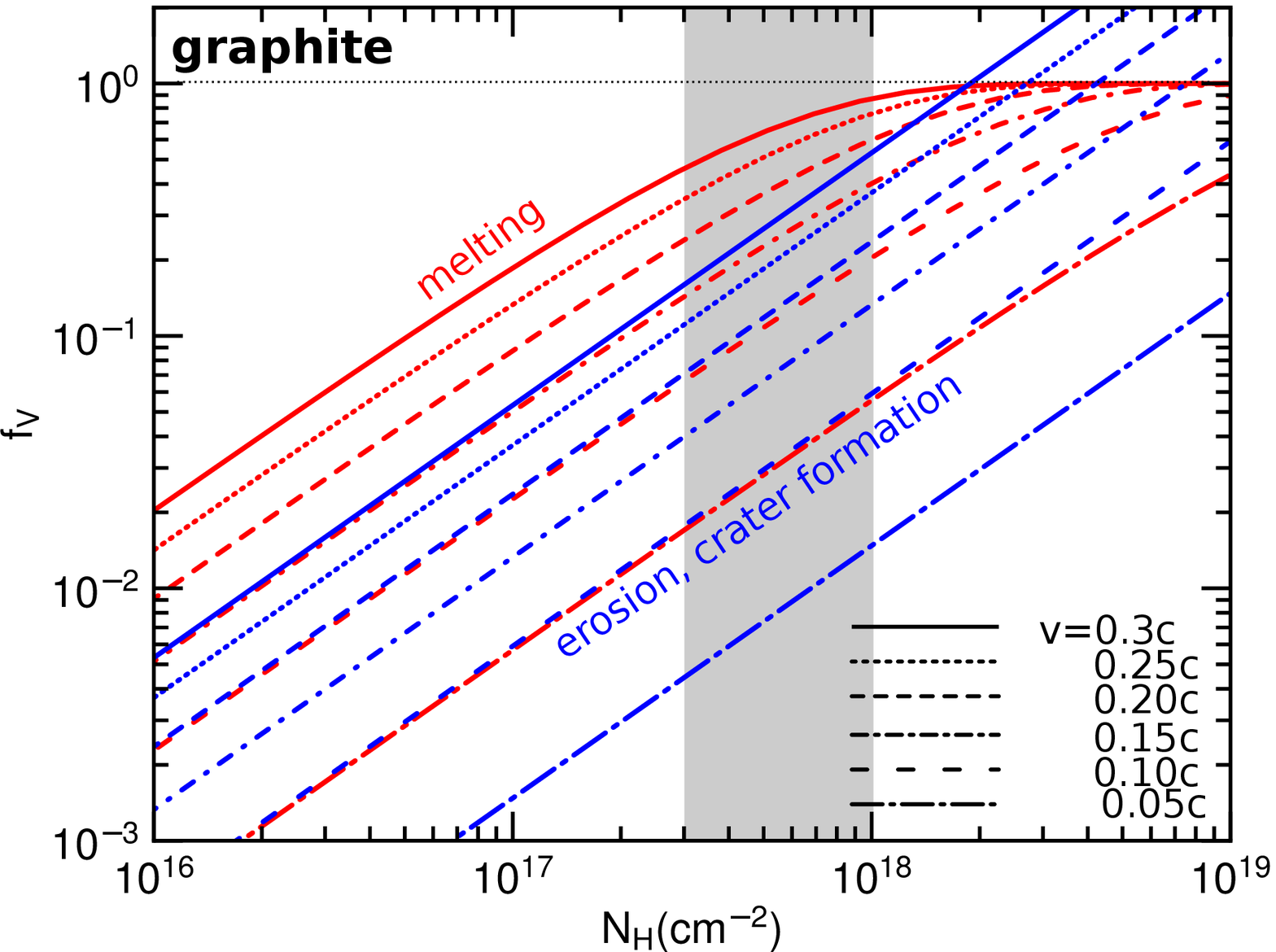}
\caption{Fraction of the spacecraft volume eroded (red lines) and melted (blue lines) due to bombardment of interstellar dust grains computed for quartz (left panel) and graphite (right panel) materials. A thickness of the surface layer $L=1$ cm is assumed.}
\label{fig:fvol_dust}
\end{figure*}

Figure \ref{fig:fvol_dust} shows the fraction of the spacecraft volume, $f_{V}$, eroded by dust bombardment (red lines) for quartz (left panel) and graphite (right panel), {assuming the spacecraft surface of $L=1\cm$ thick}. {The value of $f_{V}$ increases rapidly with increasing $v$, in contrast to the fraction of damaged surface area $f_{S}$, consistent with the fact that the faster dust grain can penetrate deeper into the spacecraft}. For $v=0.1-0.2c$, the spacecraft surface will be eroded up to $30\%$ of its volume by the time the spacecraft reaches $\alpha$ Centauri, passing the ISM with the column density $N_{\rm H,obs}\sim 3\times 10^{17}-10^{18}\cm^{-2}$. The fraction of the volume melted by dust bombardment is presented by the blue lines. For $v=0.1-0.2c$, the entire spacecraft of quartz material may be melted {to a depth of $L=1$ cm}, but melting for graphite is much less efficient, as expected.

\subsubsection{Evaporation by whole target heating}
When the size of the dust grain is sufficiently large, it can result in complete destruction of the spacecraft after a single collision. Indeed, due to the macroscopic size of the projectile grain, atoms in the grain interior interact with less target atoms than the outer ones, allowing them to penetrate deeper into the target. As a result, the complete destruction of the spacecraft is perhaps possible.

{The critical size $a_{d,c}$ of the grain required for complete destruction of the spacecraft can be evaluated by setting $dV$ in Equation (\ref{eq:dV_crater}) to the spacecraft volume $LA=LH^{2}$:}
\bea
\frac{a_{d,c}}{H} &=&\left(\frac{6n_{s}U_{0}}{4\pi n_{d}\epsilon_{V} mv^{2}}\right)^{1/3}\left(\frac{L}{H}\right)^{1/3},\nonumber\\~~~\label{eq:ac_aT}
&\simeq& 0.002\epsilon_{V}\left(\frac{12n_{s}}{\tilde{M}n_{d}}\right)^{1/3}\left(\frac{U_{0}}{6\eV}\right)^{1/3}\left(\frac{0.2c}{v}\right)^{2/3}\left(\frac{L}{H}\right)^{1/3},~~~~~
\ena
where $\tilde{M}$ is the average atomic mass of the dust.

{Figure \ref{fig:fm-evap} shows the values of $a_{d,c}$ as a function of the spacecraft length $L$ for $H=0.1$ cm and $0.3$ cm. For one gram mass spacecraft (e.g., $L=5\cm$ and $H=0.3\cm$ with density $\rho\sim 2.2\g\cm^{-3}$) moving at $v=0.2c$, a very big grain of $a_{d,c}\sim 15\mum$ may destroy it after a single collision. Smaller spacecrafts of $L=1\cm$ and $H=0.1\cm$ are evaporated by large grains of $a_{d,c}\sim 4\mum$, at the same speed. }

\begin{figure}
%\centering
\includegraphics[width=0.4\textwidth]{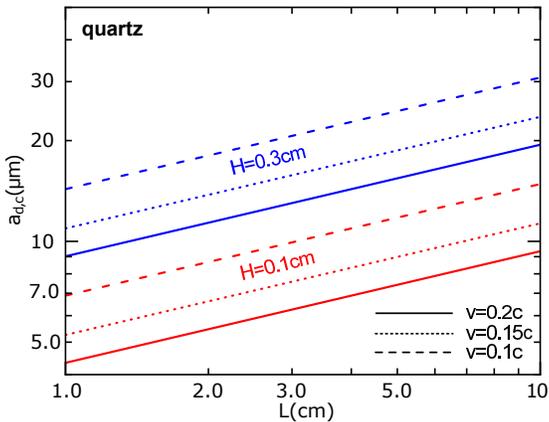}
\caption{Critical size of the projectile grain that would result in the complete evaporation of the spacecraft versus the spacecraft length $L$ for $H=0.1$ cm and $0.3$ cm.}
\label{fig:fm-evap}
\end{figure}

The total collision rate for the complete evaporation is obtained by integrating $R_{\rm coll}$ over the grain size distribution from $a_{d,c}$ to $a_{\max}$:
\bea
R_{\coll}= \sum_{j}\int_{a_{d,c}}^{a_{\max}}Av \frac{dn_{j}}{da} da.~~~~
\ena

Assuming a constant gas-to-dust mass ratio, the column density of gas swept by the spacecraft before its complete destruction is $N_{\coll}=n_{\H}v R_{\coll}^{-1}$. The chance of colliding with destructive grains of $15\mum$ is less than unity up to a huge gas column $N_{\coll}\sim 10^{68}\cm^{-2}$, assuming $A=0.3\times 0.3\cm^{2}$ and $v=0.2c$. 

\section{Heating of Spacecraft and its Damage}\label{sec:heat}
\subsection{Heating, Cooling and Equilibrium Temperature}
In Section \ref{sec:ISMgas} we show that transient heating by heavy atoms can produce damage track of a few nanometers, while light atoms convert their energy into heating the entire spacecraft without inflicting any significant damage. In this section, we will evaluate the equilibrium temperature of the spacecraft due to collisional heating and radiative heating by absorption of interstellar starlight.

Collisional heating is dominated by light elements (H and He) because they contain most of the gas mass. Thus, the rate of collisional heating to the spacecraft surface can be written as
\bea
\frac{dE_{\rm h, \coll}}{dt} &=& \sum_{i}x_{i} n_{\H}vA\left(\frac{m_{i}v^{2}}{2}\right)\simeq  \frac{1.4n_{\H}m_{\H}v^{3}A}{2},\nonumber\\
&\simeq &2.5\times 10^{5}n_{\H}A\left(\frac{v}{0.2c}\right)^{3} \erg\s^{-1},
\ena
where the factor of $1.4$ accounts for the abundance and mass of He relative to H, and the minor contribution of heavier atoms is ignored.

The rate of radiative heating by the interstellar radiation (ISRF) of spectral energy density $u(\nu)$ is given by
\bea
\frac{dE_{\rm h, rad}}{dt}=\int d\nu cu(\nu)Q_{\abs,\nu}A=cu_{\rad}A.
\label{eq:dEhdt_iso}
\ena
where  $u_{\rm rad}=\int u(\nu) d\nu$, and the absorption efficiency is denoted by $Q_{abs,\nu}$.

For the ISRF with $u_{\rm rad}=8.64\times 10^{-13}\erg\cm^{-3}$ (\citealt{1983A&A...128..212M}), we have $dE_{\rm h,rad}/dt \sim0.026A\langle Q_{\abs}\rangle \erg\s^{-1}$.  Therefore, for a relativistic spacecraft, the radiative heating by the ISRF is negligible compared to collisional heating by interstellar gas. 

The front surface of the spacecraft also emit thermal radiation, which results in radiative cooling at rate:
\bea
\frac{dE_{\rm c, rad}}{dt}=\int d\nu AQ_{\abs, \nu}B_{\nu}(T)
=A\langle Q_{\abs}\rangle_{T} \sigma T^{4},~~~~
\label{eq:dEcdt}
\ena
where
\bea
\langle Q_{\abs}\rangle_{T}=\frac{\int d\nu Q_{\abs,\nu}B_{\nu}(T)}{\int d\nu B_{\nu}(T)} \label{eq:Qabsavg}
\ena
 is the Planck-averaged emission efficiency. The emission efficiency $\langle Q_{\abs}\rangle_{T}=1$. Here, the Doppler shift correction is negligible for $v\le 0.2c$.

The surface temperature can be estimated by balancing the collisional heating and the radiative cooling i.e.
\bea
T_{\rm eq}&\simeq& \left(\frac{1.4 n_{\H}m_{\H}v^{3}}{2\sigma} \right)^{1/4},\nonumber\\
&\simeq& 258\left(\frac{n_{\H}}{1.0\cm^{-3}}\right)^{1/4}\left(\frac{v}{0.2c}\right)^{3/4}\K.\label{eq:Teq}
\ena

\begin{figure}
\includegraphics[width=0.45\textwidth]{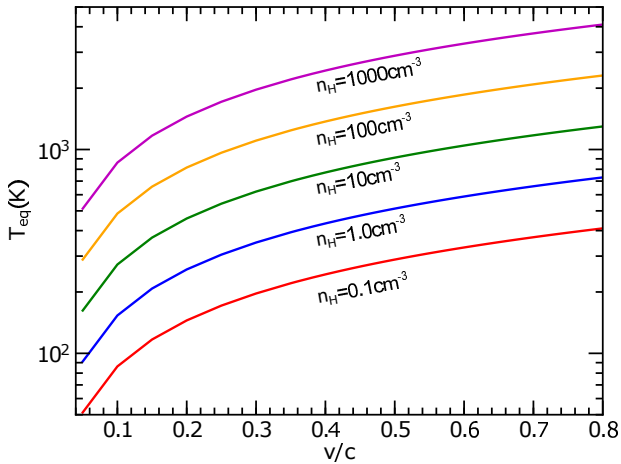}
\caption{Surface temperature of the spacecraft vs. its speed due to collisional heating by interstellar gas of various densities.}
\label{fig:Teq}
\end{figure}

Figure \ref{fig:Teq} shows the surface temperature of the spacecraft at different speeds in the interstellar gas of various densities of $n_{\H}=0.1-10^{3}\cm^{-3}$. The average density along the line of sight to $\alpha$ Centauri is $\bar{n}_{\H}=N_{\rm H,obs}/1.34 {\rm pc}\simeq 0.07-0.24\cm^{-3}$.
Therefore, unless there exist clumps of density $n_{\H}\ge 10^{3}\cm^{-3}$ along the spacecraft journey toward $\alpha$ Centauri, the diffuse ISM of $n_{\H}\le 10\cm^{-3}$ only heats the spacecraft to surface temperature of $T<T_{m}$, which is insufficient to cause serious damage such as melting. 

\subsection{Heat transfer and temperature profile in the spacecraft}
When the spacecraft surface is heated to a high temperature, such as by passing a very dense clump, heat conduction from the surface toward the inner spacecraft must be studied in order to assess potential damage to electronic components by overheating.

Let $T_{0}$ be the surface temperature. At the surface layer, energy conservation gives:
\bea
dE_{\rm coll}/dt + dE_{\rm rad}/dt =\sigma T_{0}^{4}\langle Q_{\rm abs}\rangle_{T}A + \dot{Q}_{\rm cd}A,\label{eq:bal}
\ena
where $\dot{Q}_{\rm cd}$ is the heat flux transported from the surface to the inner layer through heat conduction. 

In the case of steady heat conduction (i.e., constant heat flux), the temperature at depth $x$ from the surface can be described by the heat conduction equation:
\bea
-\frac{\kappa dT}{dx}=\dot{Q}_{\rm cd},\label{eq:qflux}
\ena
where the thermal conductivity coefficient $\kappa$ is given by
\bea
\kappa = \alpha\rho c_{p}(T),\label{eq:kappa}
\ena
where $c_{p}(T)$ is the specific heat capacity at temperature, and $\alpha$ is the thermal diffusivity. 

For the case of low heat conductivity, the value of $T_{0}$ can be directly obtained from Equation (\ref{eq:bal}), which provides $T_{0} \sim T_{\rm eq}$ with $T_{\rm eq}$ given by Equation (\ref{eq:Teq}). 

The temperature at depth $x$ from the surface can be obtained by the following equation:
\bea
\int_{T_{0}}^{T}\alpha \rho c_{p}(T) dT = \dot{Q}_{cd} x,\label{eq:kappa_T}
\ena
where the explicit dependence of $\kappa$ on $T$ is accounted for. 

We numerically solve Equation (\ref{eq:kappa_T}) using the function of $c_{p}(T)$ for quartz and graphite from \cite{2001ApJ...551..807D}. We consider the range of diffusivity $\alpha=0.01-0.05\cm^{2}\s^{-1}$ for quartz, and $\alpha=0.1-2 \cm^{2}\s^{-1}$ for graphite.

Figure \ref{fig:Tprof} shows the derived temperature profile for quartz and graphite in the interstellar gas of various number density $n_{\H}=0.1,1,~10\cm^{-3}$. The lower diffusivity of quartz results in slow heat conduction, yielding a larger temperature difference over a depth of $10$cm. 
Standard graphite has high thermal conductivity of $\alpha=2\cm^{2}\s^{-1}$. Therefore, the temperature is slightly different over a depth of $10\cm$. 

\begin{figure*}
\includegraphics[width=0.33\textwidth]{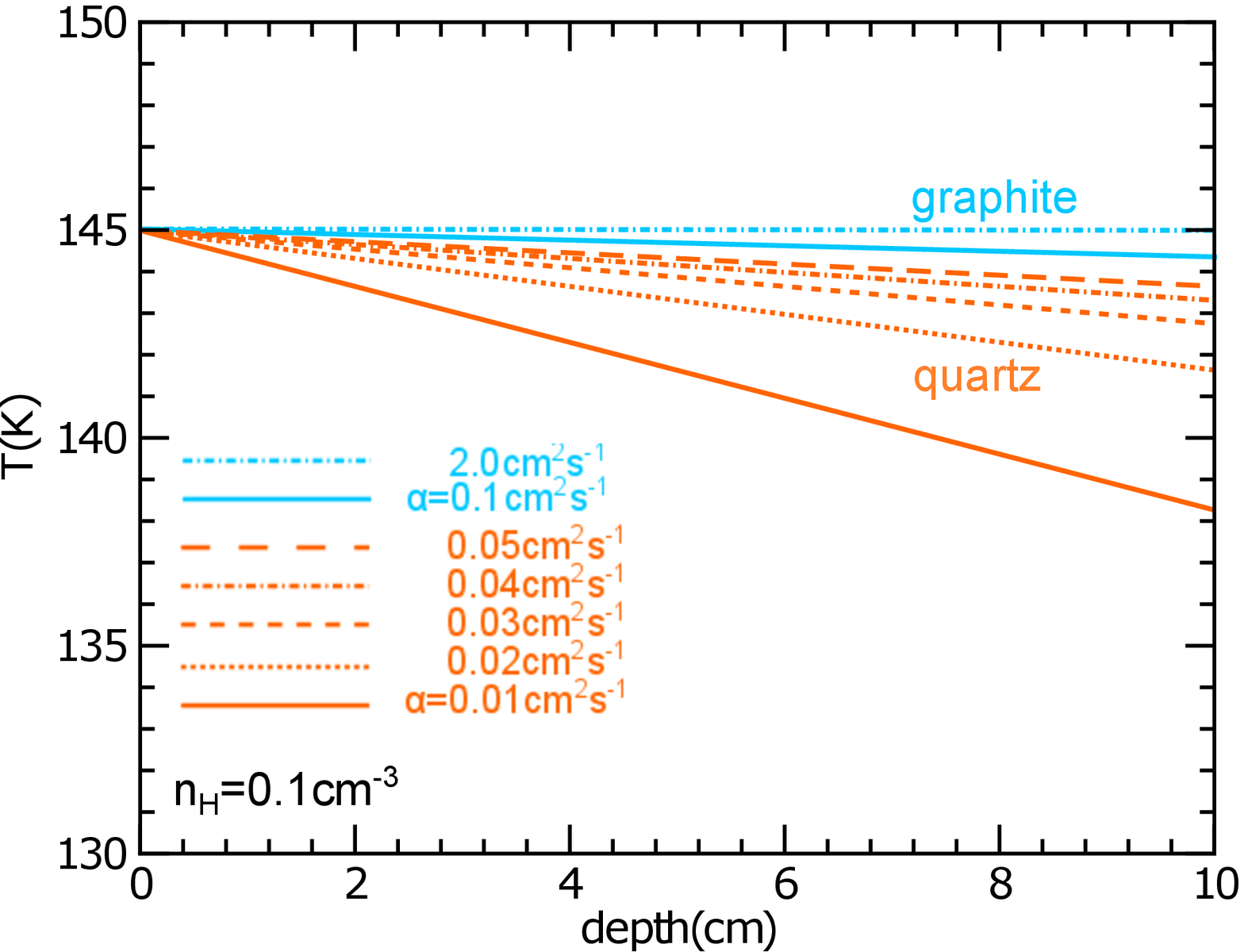}
\includegraphics[width=0.33\textwidth]{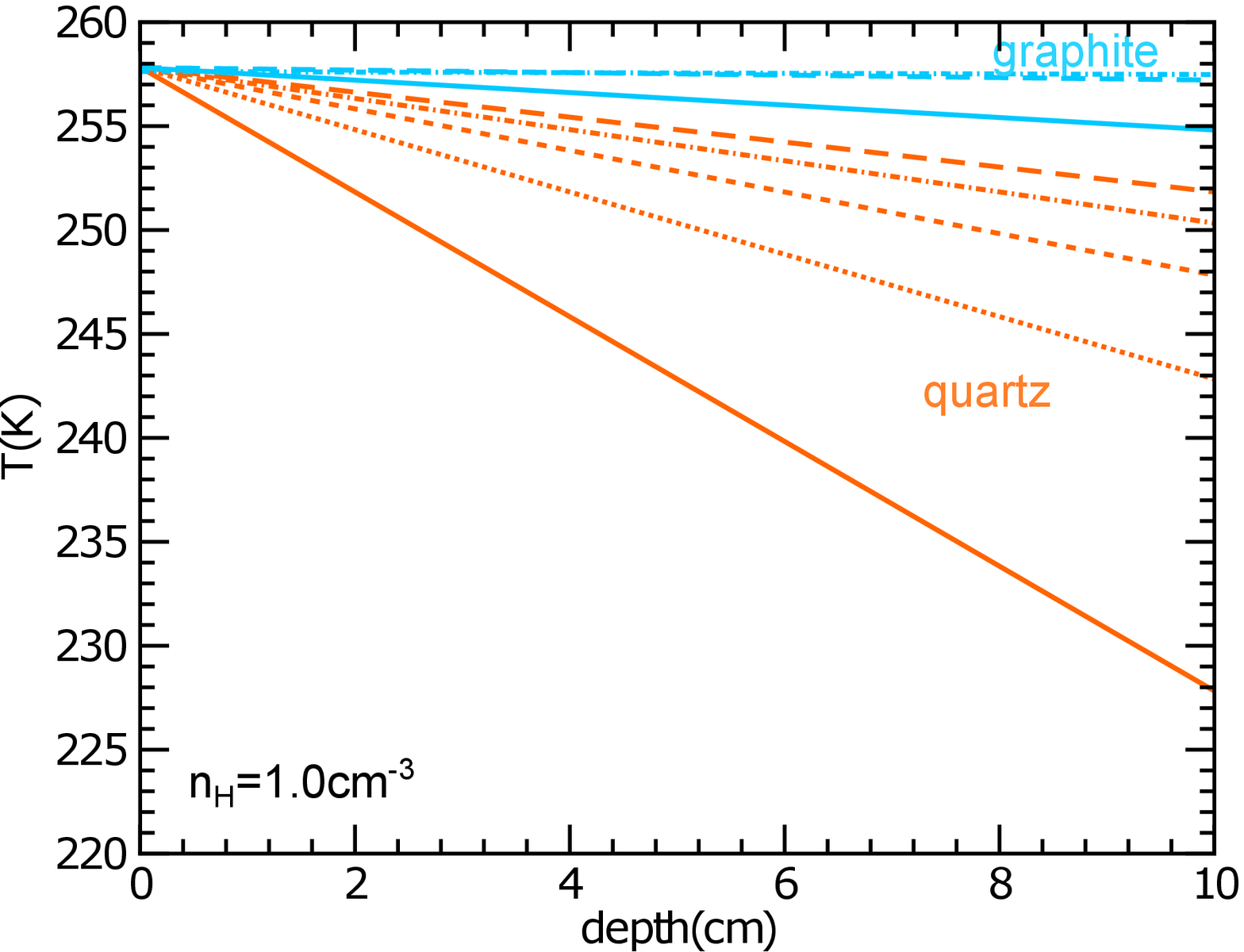}
\includegraphics[width=0.33\textwidth]{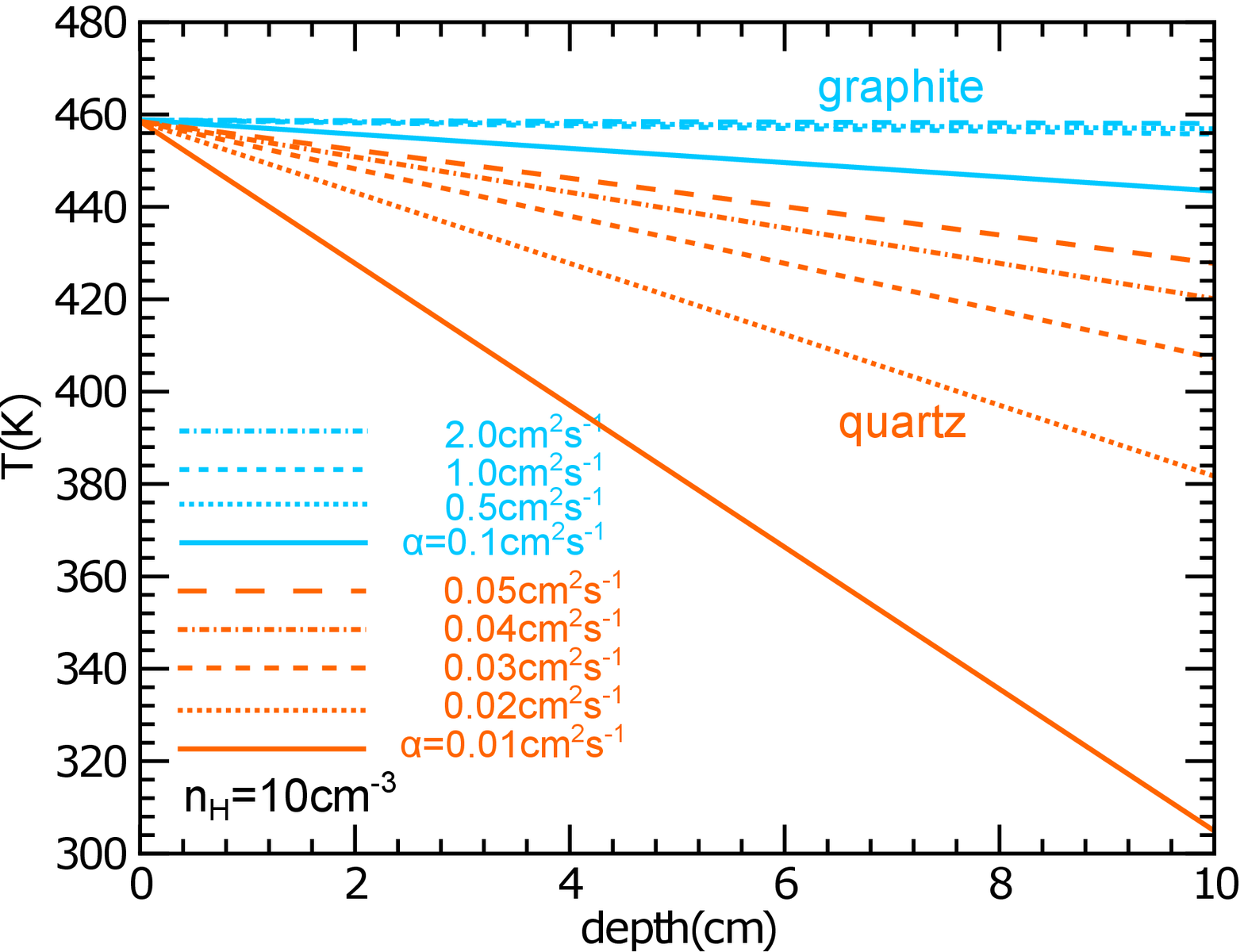}
\caption{Temperature profile as function of the depth inside the spacecraft's material for different thermal diffusivity values in quartz ($\alpha=0.01-0.05 \cm^{2}\s^{-1}$) and graphite ($\alpha=0.1-2 \cm^{2}\s^{-1}$). Three values of gas density $n_{\H}=0.1, 1.0,10\cm^{-3}$ and a spacecraft speed $v=0.2c$ are considered.}
\label{fig:Tprof}
\end{figure*}

Our results from the sections above suggest that a material with low thermal conductivity is advantageous in avoiding the heating problem if the spacecraft encounters gas clouds of enhanced density. Spacecrafts made of high thermal conductivity material may suffer damage in these situations. But if there is no dense clumps along its journey, the spacecraft has no serious problem with heating from the diffuse interstellar gas.

%\newpage
\section{Discussion}\label{sec:dis}
\subsection{Damage of spacecraft due to interstellar gas and dust}
\subsubsection{Our main results}
We have quantified the damage to a spacecraft of similar specifications as the proposed Breakthrough Starshot spacecraft due to collisions with interstellar gas and dust.  We considered  two types of materials: quartz and graphite. The major effect of collisions with gas atoms is the damage of the surface area due to track formation. This type of damage mostly results the reduction in the strength of the material structure. We find that damage by energetic ions is most important for quartz, whereas graphite material of high conductivity is damaged only when moving at a speed $v\le 0.15c$. At larger speeds, graphite is not damaged considerably because of the decrease in the ion energy loss with increasing $v$.

Interstellar dust bombardment induces transient spot heating, which can result in the sudden evaporation of atoms from the surface, producing craters on the spacecraft surface. Crater formation by interstellar dust is important for both quartz and graphite composition, although the evaporation depends on the binding energy which is slightly different for these two materials. In addition, dust bombardment also results in sudden melting of the surface. This melting process does not erode the surface but modifies its structure and may cause electronic devices to malfunction.

Figure \ref{fig:Ldm} shows the thickness of the surface layer damaged by dust bombardment and gas bombardment as a function of $N_{\H}$ for the different speeds. The value of $L_{\rm dm}$ is determined by the thickness $L$ at which $f_{V,\rm evap}=1$ for evaporation (Equation \ref{eq:fvol_dust}) and $f_{V,m}=0.9$ for melting (Equation \ref{eq:fvol_dust_melt_cor}, dust bombardment) or track formation (Equation \ref{eq:fV}) gas bombardment). 
At speed $v=0.2c$, the entire surface can be evaporated to the depth of $\sim 0.5$ mm (0.7mm) whereas melting can damage to a larger depth of 3 mm (1 mm) for quartz (graphite) after the spacecraft has swept a column density of $N_{\H}\sim 3\times 10^{17}\cm^{-2}$. Since the expected gas column towards $\alpha$ Centauri is $N_{\rm H,obs}\sim 3\times 10^{17}-10^{18}\cm^{-2}$ (shaded area in Figure \ref{fig:Ldm}), the expected damage is up to three times larger.

\begin{figure*}
\centering
\includegraphics[width=0.45\textwidth]{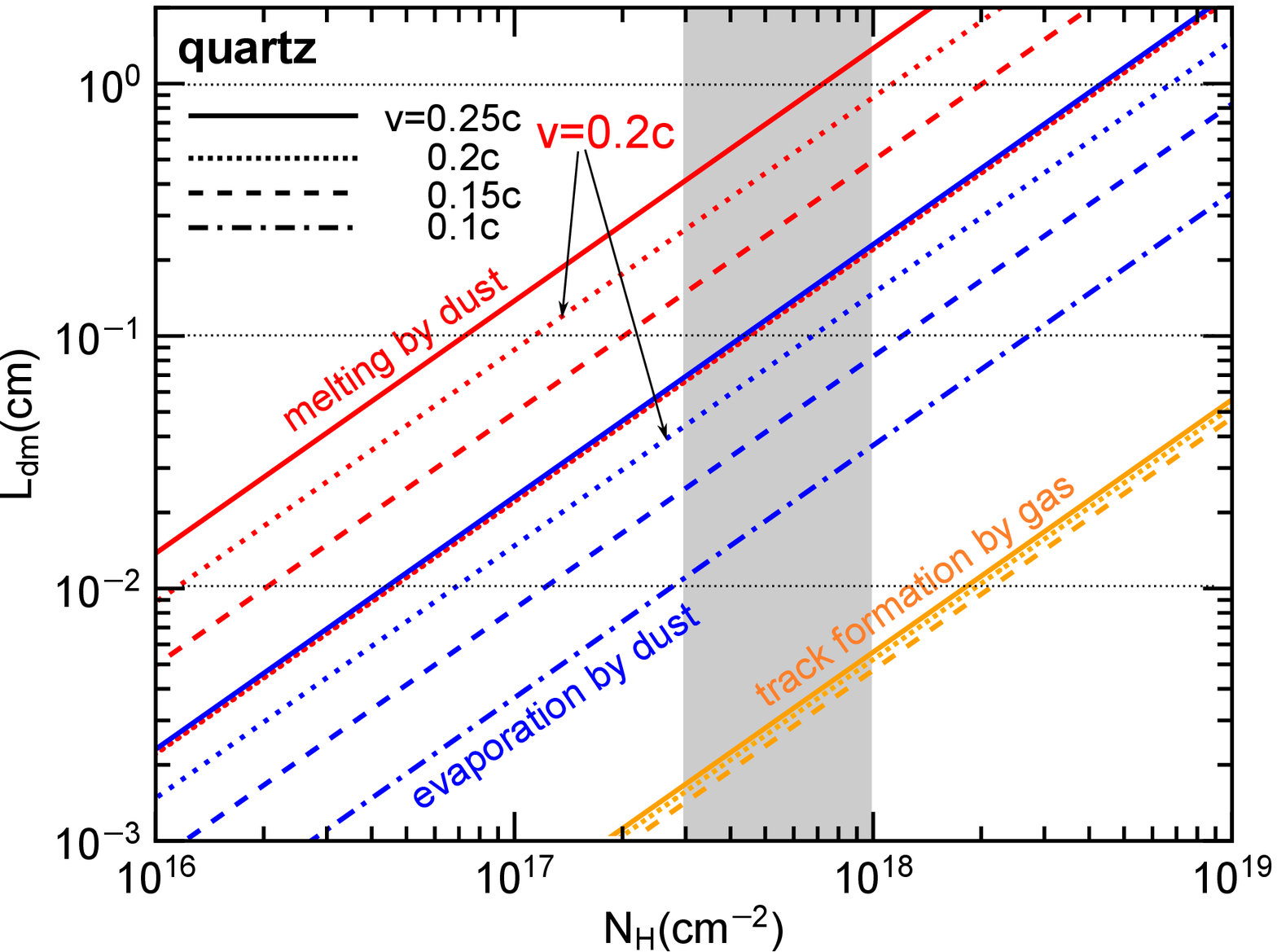}
\includegraphics[width=0.45\textwidth]{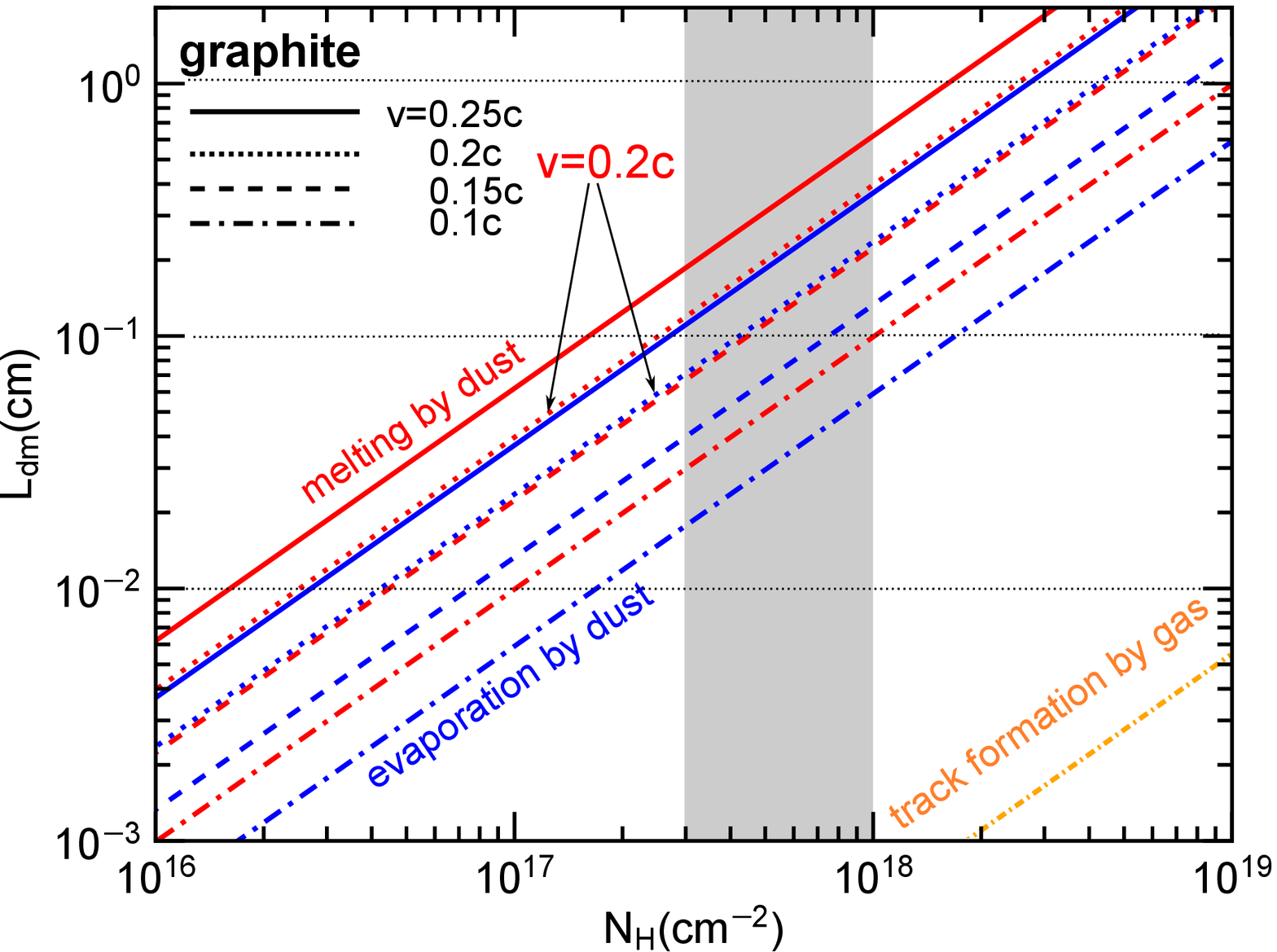}
\caption{Thickness of surface damaged by dust bombardment and gas bombardment for quartz (left) and graphite (right). Melting for graphite is less efficient than for quartz because of its high melting temperature.}
\label{fig:Ldm}
\end{figure*}

Figure \ref{fig:Ldm} shows that explosive evaporation (melting) by interstellar dust is at least one order (two orders) of magnitude more efficient than the damage by interstellar gas. While containing $\sim 1$ percent of mass, dust is composed of only heavy elements. Moreover, an individual heavy atom of atomic mass $M<50$ cannot produce damage track at $v\sim 0.2c$ because of low energy loss $dE/dx$, but a dust grain containing $\sim 10^{10}$ of such heavy atoms can deposit a huge amount of energy to a larger cylinder that induces sudden evaporation of a spot on the spacecraft.
 
Finally, spacecrafts of one gram mass (i.e., $L=5\cm, H=0.3\cm$) can be completely destroyed by interstellar dust grains larger than $15\mum$ after a single collision. However, this very big grain population is extremely rare in the ISM, and along the entire trajectory to $\alpha$ Centauri, the chance of encountering one such grain is negligible of $\sim 10^{18}/10^{68}=10^{-50}$. 

\subsubsection{Uncertainty in the abundance of very big grains in the ISM}
{Our estimates of the damage by interstellar dust, including the extremely low chance of hitting very big grains ($\sim 10^{-50}$) are obtained by using the standard size distribution of interstellar grains from \cite{2001ApJ...548..296W}, where the abundance of very big grains of size $a>1\mum$ is very low. 

Nevertheless, the grain size distribution in the local interstellar medium may be different from that of the average Galactic ISM (Eq. \ref{eq:dnda}). For instance, the analysis of data from Ulysses and Galileo spacecrafts in \cite{1999ApJ...525..492F} shows that the local ISM contains a large amount of micron-sized grains, with the power law has upper limit up to $\sim 3\mum$. In particular, radar automatic surveys have detected interstellar meteoroids between $10-30\mum$ \citep{2000JGR...10510353B}. Various measurements (see \citealt{2012ApJ...745..161M}, Figure 1) show a flux of interstellar particles (IP) as a function of the particle mass $m$: $f_{\rm IP}\sim 10^{5}(10^{-12}\g/m)^{-1.1} {\rm km}^{-2}h\sim 0.09 (10^{-12}\g/m)^{-1.1}\cm^{-2}\yr^{-1}$. With these measurements, the chance of encountering a very big grain of size above $10\mum$ is $7.99\times 10^{-8}$ over a journey of 20 years. 

\cite{Poppe:2016gr} presents a model of interplanetary dust that reproduces the in-situ data. According to this study, the number density of very big particles of size $a \ge 5\mum$ at heliodistance $d$ is $n_{\rm IP}\sim 10^{-18}\cm^{-3}(d/70\AU)^{-\eta}$ where $\eta>0.5$. Integrating over the entire journey to distance $d$, we can estimate the number of the $a \ge 5\mum$ particles that the spacecraft can encounter is $N_{\rm IP}\sim  0.13 (d/1.3\pc)^{-\eta/0.5}(A/1\cm^{2})$. For a cross-section $A=0.3\times 0.3\cm^{-2}$, we get $N_{\rm IP}\sim 0.01(d/1.3\pc)^{-\eta/0.5}$, thus it is unlikely that the spacecraft will be destroyed by collisions with very big particles ($a \ge 5\mum$). Moreover, reducing $A$ can reduce the chance of colliding with such big particles. We note that the Poppe's model was developed for dust in elliptic plane whereas the direction to $\alpha$ Centauri is out of this plane. Therefore, the column density of such big grains is expected to be much lower than $N_{\rm IP}$, and the chance of spacecraft destruction via this process is much smaller.}

%{\bf With all the uncertainty in the abundance of very big grains in the ISM, we expect that the chance of meeting risky grains is within the range determined by our calculations using the WD01 model and the estimates above using the current data.}

\subsection{Possible ways to protect the spacecraft}

Our study has identified the risk of damage from both gas bombardment and dust collisions to the spacecraft and so we now consider the ways of mitigating those risks. 
The first obvious step towards protecting the spacecraft from bombardment of large dust particles is to minimize the cross sectional area {(see also \citealt{2016arXiv160401356L})}.
This is because the rate of hits scales as the cross sectional area as well as the speed.  
The surface does not play a role in protecting against gas bombardment since, after the spacecraft traverses 1 pc, 
100\% of its surface is eroded away.  This damage can be mitigated with protective layering made of high conducting material such as graphite. 
The concern with protective layers for the Starshot mission is the weight of the spacecraft as the entire system should be on the gram-scale. 
Therefore, we recommend both to add a protective layer and to minimize the spacecraft
incident surface area (i.e., to avoid large grain hits as well as to minimize the mass of the protective layer).}

To prevent damage by gas bombardment, the spacecraft can be protected with a thin layer of 0.01 mm made of highly conducting material, such as graphite or beryllium. 
For dust bombardment, the crater formation is insensitive to the material because it is determined only by the binding energy and the total thickness of protective shielding. Our results suggest that a shield made of graphite of $\sim$ 1-3 mm thickness will be required to prevent the melting by dust bombardment. {A shield made of Be of several mm thick is suggested by \cite{2016arXiv160401356L}.}    

{For a thin lightsail, gas atoms essentially pass through the sail without damage (see Section \ref{sec:ISMgas}). Dust grains will likely produce a number of punch with size comparable to the dust grain radius (\citealt{2015JBIS...68..205E}).}
In order to protect the lightsail from dust/gas bombardment the lightsail should be folded and retracted behind the protective material in a needle-like configuration.  Another solution may be to put the lightsail behind the protective material but in front of the Starchip, to further protect the more sensitive electronic equipment.  Retracting the lightsail behind the protective coating will also reduce mechanical torque on the spacecraft due to surface irregularities. 

{A thin foil may be put in front of the spacecraft at some distance, such as dust grains will be exploded by Coulomb explosions before hitting the spacecraft (Jim Early, private communication). The foil must be sufficiently thick to slow down the dust ions so that they will cause minor damage to the spacecraft via sputtering effect.}
%The issue with protective layers for the Starshot mission is the weight of the craft as the entire device should be on gram-scales. 

%After launch and during the cruse phase of the mission, the lightsail can be folded down to be in front of the Starchip in order to protect the electronic equipment and to protect the sail itself. But this would compromise the sail

\subsection{Deflection of dust particles}
We found that larger interstellar dust grains play a dominant role in the damage of a relativistic spacecraft. Thus, it is crucial to deflect them from the path of spacecraft. Below, we discuss two potential ways to deflect interstellar dust. Dust particles can be optically detected ahead of the spacecraft and deflected or destroyed.

\subsubsection{Electric deflection}
Relativistic spacecrafts accumulate positive charge through photoelectric emission while moving in the ISM.
Since large interstellar grains have positive charge (see \citealt{2011piim.book.....D}), they can be deflected by strong electric field of the spacecraft. 

Let $Z_{\rm sp}e$ be the positive charge that the spacecraft has accumulated in the ISM. The closest distance that a dust grain of positive charge $Z_{d}e$ can approach the spacecraft is determined by
\bea
\frac{m_{d}v^{2}}{2}=\frac{Z_{d}Z_{\rm sp}e^{2}}{r},
\ena
which yields
\bea
r_{\min} =\frac{2Z_{d}Z_{\rm sp}e^{2}}{m_{d}v^{2}}.
\ena
Plugging $Z_{\rm sp}=Z_{\rm sp, max}=7.5\times 10^{14} a_{\rm sp}^{2}$ with $a_{\rm sp}^{3}=3LH^{2}/4\pi$ being the effective size of the spacecraft (see Eq. \ref{eq:Zmax}) and $m_{d}=1.25\times 10^{-14} a_{-5}^{3}\g$, we obtain
\bea
r_{\min}\simeq 2.3\times 10^{-8} a_{\rm sp}^{2}a_{-5}^{-3}(Z_{d}/30) (v/0.2c)^{-2}~ \cm.
\ena

Therefore, even if we can charge the dust grain to $Z_{d,\max}=7.5\times 10^{4}a_{-5}^{2}$, it still cannot help to deflect the dust grains via Coulomb repulsion because $r_{\min}$ much smaller than the spacecraft dimension $a_{\rm sp} \sim 1$cm.

\subsubsection{Radiation pressure deflection}
{Scattering and absorption of radiation from the spacecraft can also accelerate interstellar dust grains. The {\it deflecting} force that a laser beam can apply to the grain is proportional to $\sin\theta$, where $\theta$ is the angle that the laser beam makes to the spacecraft trajectory. 
The radiation pressure force that repels the dust grain out of the spacecraft trajectory is given by, 
$F_{\rm rad}={P \sin\theta}/{c}$, where
$P$ is the laser power, and perfect absorption is assumed. 
For a small spacecraft the angle $\theta$ is small, which would suggest that it is better to use two spacecrafts moving along the same line, with one of them cleaning the way to the other. 

%Let assume that we can keep the laser beam pointing to the dust grain. Then,
%to successfully deflect the grain, the perpendicular displacement of the dust grain must be larger than the spacecraft width $W$ before the the spacecraft hits the dust grain. Therefore, the power required is given by
%\bea
%P &\sim & \frac{2m_{gr}v_{sp}^{2}W}{\sin\theta d^{2}}\simeq 0.9\left(\frac{v_{sp}}{0.2c}\right)^{2}\left(\frac{W}{0.3\cm}\right)\nonumber\\
%&\times&\left(\frac{a}{10\mum}\right)^{3}\left(\frac{100\cm}{d}\right)^{2}\frac{1}{\sin\theta}~ {\rm Watt}.
%\ena

To enhance the force impact on the dust grain, one can evaporate part of dust grain with the evaporated particles acting as a rocket jet for the purpose of deflection. The volatiles on the particle (e.g., ices) can be easily evaporated. Indeed, this can be used to mitigate the damage from big particles where it is possible to heat the particle on one side.}
 
\subsection{Effect of interstellar magnetic field on the spacecraft trajectory}
Since the spacecraft are positively charged, its trajectory may be be affected by the interstellar magnetic field $\Bv$. The gyroradius of a charged spacecraft moving across $\Bv$ with perpendicular velocity $v_{\perp}$ is
\bea
r_{g} = \frac{m_{\rm sp}cv_{\perp}}{Z_{\rm sp}eB}\simeq 150\left(\frac{v_{\perp}}{0.2c}\right)\frac{\hat{\rho}a_{\rm sp}}{\hat{B}}\frac{Z_{\rm sp,\max}}{Z_{\rm sp}} {~\rm pc},\label{eq:rgLarmor}
\ena
where $Z_{\rm sp}e$ is the equilibrium charge, {and $\hat{\rho}=\rho/3\g\cm^{-3}$ with $\rho$ being the mass density of the spacecraft, $\hat{B}=B/10\mu G$ with B the magnetic field strength.}

We find that the Larmor radius is about 150 pc for a maximally charged spacecraft. Therefore, the effect of magnetic fields is negligible.  

\subsection{Thermal energy battery}
We have found that the spacecraft can be dominantly heated by interstellar gas. For an average density $\bar{n}_{\H}\sim 0.1\cm^{-3}$, the temperature is about $\sim 280 \K$ at speed $v=0.2c$. If there exists some dense clumps of density $n_{\H}\sim 100\cm^{-3}$ along the journey, the spacecraft surface can be heated to $T\sim 800$ K. This uniform temperature is insufficient to damage the spacecraft, but its heat may be used to power electronic devices. 

A potential method for energy storage is to use thermal battery. For this purpose, the temperature difference must be sufficiently large. We find that materials of low conductivity (e.g., quartz) can produce larger temperature difference. Similarly, elongated spacecraft may be advantageous for having a large temperature difference, as shown in Figure \ref{fig:Tprof}.

\subsection{Comparison to other studies}
{\cite{2016arXiv160401356L} discussed the risk from interstellar dust by estimating the total number of collisions with interstellar dust by the time the spacecraft reaches $\alpha$ Centauri for several shapes of the spacecraft, but did not quantify the consequence of dust collisions to the spacecraft. In this paper, we have investigated in detail the consequence of dust collisions by applying the microphysics of collisions of energetic particles on solid. Although the energy transfer of the dust grains is rather small compared to the spacecraft kinetic energy, as pointed out by \cite{2016arXiv160401356L}, we have found that such energy from dust collisions can heat the spacecraft surface to high temperatures, resulting in melting and craters. We have also studied the damage by energetic gas atoms and found that heavy ions, such as iron, can damage the spacecraft surface to a few mm depth by means of track formation.}

%\subsection{Implications for other astrophysical events}
% \cite{2015ApJ...806..124G} suggested that stars originally bound to massive black holes can be disrupted and are moving with relativistic velocities. This similar mechanism perhaps produce relativistic meteorites. The motion of such relativistic meteorites through some dense cloud of the ISM would produce flash of thermal radiation, which may be observable from ground.  

\section{Summary}\label{sec:summ}
We have investigated in detail the interaction of a relativistic spacecraft with gas atoms and dust grains in the interstellar medium on the journey toward the nearest star system, $\alpha$ Centauri. The principal results are summarized as follows:

\begin{itemize}
\item [1] We find that heavy atoms in the interstellar gas can transiently produce damage tracks of several nanometers radius in the spacecraft, which lead to the modification of the material structure. Through this effect, interstellar gas can damage the spacecraft surface to a depth of $\sim 0.1$ mm for quartz composition, after the spacecraft sweeps a gas column density $N_{\H}\sim 2\times 10^{18}\cm^{-2}$. If the spacecraft is made of highly conductivity material, such as graphite, damage by heavy gas atoms can be prevented by quickly transferring their energy throughout the spacecraft and therefore averting the track formation. 

\item[2] Interstellar dust can produce numerous craters on the spacecraft surface as a result of explosive evaporation following each dust grain encounter. This effect can erode the entire surface of the spacecraft to a thickness of $\sim $ 0.5 mm after it has swept a gas column of $N_{\H}\sim 3\times 10^{17}\cm^{-2}$ for $v\sim 0.2c$.  This column density is lower than the measured column of $N_{\rm H,obs}\sim 3\times 10^{17}-10^{18}\cm^{-2}$ towards $\alpha$ Centauri. Dust bombardment also induces melting of the surface layer and modify its structure, which is more efficient than explosive evaporation.

\item[3] We estimated that an encounter with a dust grain larger than 15 $\mu$m will completely destroy gram-scale spacecrafts. {Given the low abundance of very big grains in the ISM, their effect is likely to be unimportant.}

\item[4] We calculated the equilibrium temperature of the spacecraft due to heating by collisions with gas atoms (dominated by light elements, H and He) and interstellar radiation field. For the local diffuse medium of density $n_{\H}\le 10\cm^{-3}$, the temperature is insufficient to induce any melting.

\item[5] We have identified several ways to protect the spacecraft, a needle-like configuration as well as materials suitable for the lightsail and protective layers using the obtained quantitative estimates. {Edge treatment is discussed in \cite{2016arXiv160401356L}.}

\end{itemize}
\acknowledgments
{We thank Jim Early for the suggestion of exploding dust by a foil before its arrival and useful comments. We also thank an anonymous referee for useful comments and suggestions.}
This work was supported in part by a Starshot grant from the Breakthrough Prize Foundation to Harvard University (with A. Loeb as the PI). T.H. acknowledges the support from the Natural Sciences and Engineering Research Council of Canada (NSERC). A.Lazarian acknowledges the financial support from NASA grant NNX11AD32G. B.B. is supported by the NASA Einstein Fellowship. 

\appendix
\section{A. Dust physics}

%\subsection{A.1. Thermal Sublimation}

%\cite{1989ApJ...345..230G} investigated the sublimation of dust grains using detailed balance and derived the sublimation rate for a grain of radius $a$:
%\bea
%\frac{da}{dt}=-n^{-1/3}\nu_{0}\exp\left(\frac{-B}{kT_d}\right),\label{eq:dasdt}
%\ena
%where $n$ is the atomic number density, $B$ is the sublimation energy per atom, $\nu_0 = 2\times 10^{15} \s^{-1}$ and $B/k=68100 -20000N^{-1/3}\K $ for silicate grains,  $\nu_0 = 2\times 10^{14} \s^{-1}$ and $B/k=81200-20000N^{-1/3} \K $ for carbonaceous grains with $N$ being the total number of atoms of the grain (\citealt{1989ApJ...345..230G}; \citealt{2000ApJ...537..796W}). Here, if $T_{d}> B/k$, then the exponential terms goes to unity, which corresponds to the evaporation regime at the evaporation rate $da/dt \sim n^{-1/3}\nu_{0}$.

\subsection{A.1. Maximum charge of Starchip}
Efficient charging by photoelectric emission and collisional ionization can rapidly increase the positive charge of the Starchip, which results in an increased electric surface potential $\phi=Ze/a$ and tensile strength $\mathcal{S}=(\phi/a)^2/4\pi$. When the tensile strength exceeds the maximum limit that the material can support $\mathcal{S}_{\max}$, the grain will be disrupted by Coulomb explosions.

Setting $\mathcal{S}=\mathcal{S}_{\max}$, we can derive the maximum surface potential and charge that the Starchip still survives:
\bea
\phi_{\max}\simeq 1.06\times 10^{3}\left(\frac{\mathcal{S}_{\max}}{10^{10} {\rm dyn} \cm^{-2}}\right)^{1/2}a_{-5} {\rm V},\label{eq:phimax}\\
Z_{\max}\simeq 7.4\times 10^4 \left(\frac{\mathcal{S}_{\max}}{10^{10}{\rm dyn} \cm^{-2}}\right)^{1/2}a_{-5}^{2}.\label{eq:Zmax}
\ena

The value $\mathcal{S}_{\max}$ is uncertain due to the uncertainty in the grain composition. Experimental measurements for ideal material provide $\mathcal{S}_{\max}\sim 10^{11} {\rm dyn} \cm^{-2}$. Assuming that the spacecraft is made of the strongest material, e.g., Tungsen,  we adopt $\mathcal{S}_{\max}\sim 10^{10} {\rm dyn} \cm^{-2}$ for our numerical considerations unless stated  otherwise. 

\section{B. Heat conduction and temperature profile in the hot cylindrical track}\label{apdx:heatcond}
\subsection{Range of electrons in solid}
It is of interest to mention the the range of the electron in solid. With a kinetic energy $E_{e}$, the range of electron can be approximately given by (see \citealt{1979ApJ...231...77D})
\bea
R_{e}\simeq 118 \left(\frac{\rho}{3\g\cm^{-3}}\right)^{-0.85}\left(\frac{E_{e}}{1~\rm keV}\right)^{1.5} {\rm \AA},\label{eq:Rel}
\ena
where $\rho$ is the mass density. For $E_{e}< 1$ MeV, the energy loss is mainly through electronic excitation and ionization, whereas radiative loss through Bremsstrahlung radiation is negligible.

\subsection{Heat conduction after energy deposition by hot secondary electrons}
Below we discuss the evolution of the temperature of the hot cylindrical track following the passage of a relativistic dust grain into the solid. { The cylinder along the grain path is instantaneously supplied with an energy per length unit, $Q$. The temperature of the hot cylinder decreases with time $t$ and radius $r$ as given by}
\bea
T(r,t) = \frac{Q}{4\pi \kappa t}\exp\left(-\frac{r^{2}}{4\alpha t} \right),\label{eq:Trt}
\ena
where $\alpha$ is thermal diffusivity and $\kappa$ is the conductivity coefficient (see e.g., \citealt{1985A&A...144..147L}). For the high temperature limit, $\kappa=\alpha \rho c=3\alpha$ is constant.

At each moment, the Gaussian distribution of the temperature versus $r$ (Equation \ref{eq:Trt}) can be approximated as a rectangular profile. Thus, the instantaneous radius of the hot cylinder can be determined by the radius at which $T(r,t)=T(0,t)/2$.  Following the energy conservation, we have:
\bea
3n_{s}k \pi R_{\rm cyl}^{2}lT_{\rm cyl}(t) = Ql.\label{eq:Rcyl_T}
\ena

Sudden evaporation occurs for $T_{\rm cyl}\ge U_{0}/3k$, and melting occurs when $T_{\rm cyl}\ge T_{m}$. Therefore, evaporation and melting induces sudden damage of the spacecraft. We estimate the surface area of such a damage using Equation (\ref{eq:Rcyl_T}).

%--------------adding references-----------------------------------
%\bibliographystyle{/Users/thiemhoang/Dropbox/Papers2/apj}
% or other styles: mcbride,plain, abbrv, acm, alpha, apalike, apj
%\bibliography{/Users/thiemhoang/Dropbox/Papers2/cites_paperApJ,/Users/thiemhoang/Dropbox/Papers2/cites_Books}
%\bibliographystyle{/home/home1/cthoang/Dropbox/Papers2/apj}
%\bibliography{/home/home1/cthoang/Dropbox/Papers2/cites_paperApJ,/home/home1/cthoang/Dropbox/Papers2/cites_Books}
\bibliography{ms.bbl}

\begin{thebibliography}{35}
\expandafter\ifx\csname natexlab\endcsname\relax\def\natexlab#1{#1}\fi

\bibitem[{Baggaley(2000)}]{2000JGR...10510353B}
Baggaley, W.~J. 2000, Journal of Geophysical Research, 105, 10353

\bibitem[{{Draine}(2011)}]{2011piim.book.....D}
{Draine}, B.~T. 2011, {Physics of the Interstellar and Intergalactic Medium}
  (Princeton, NJ: Princeton Univ. Press)

\bibitem[{Draine \& Li(2001)}]{2001ApJ...551..807D}
Draine, B.~T., \& Li, A. 2001, \apj, 551, 807

\bibitem[{Draine \& Salpeter(1979)}]{1979ApJ...231...77D}
Draine, B.~T., \& Salpeter, E.~E. 1979, \apj, 231, 77

\bibitem[{Dunlop {et~al.}(1994)Dunlop, Lesueur, Legrand, Dammak, \&
  Dural}]{1994NIMPB..90..330D}
Dunlop, A., Lesueur, D., Legrand, P., Dammak, H., \& Dural, J. 1994, Nuclear
  Instruments and Methods in Physics Research Section B, 90, 330

\bibitem[{{Early} \& {London}(2015)}]{2015JBIS...68..205E}
{Early}, J.~T., \& {London}, R.~A. 2015, Journal of the British Interplanetary
  Society, 68, 205

\bibitem[{Fano(1963)}]{1963ARNPS..13....1F}
Fano, U. 1963, Annual Review of Nuclear and Particle Sciences, 13, 1

\bibitem[{Fleischer {et~al.}(1965)Fleischer, Price, \&
  Walker}]{1965JAP....36.3645F}
Fleischer, R.~L., Price, P.~B., \& Walker, R.~M. 1965, JAP, 36, 3645

\bibitem[{Frisch {et~al.}(1999)Frisch, Dorschner, Geiss, Greenberg, Gr{\"u}n,
  Landgraf, Hoppe, Jones, Kr{\"a}tschmer, Linde, Morfill, Reach, Slavin,
  Svestka, Witt, \& Zank}]{1999ApJ...525..492F}
Frisch, P.~C., Dorschner, J.~M., Geiss, J., {et~al.} 1999, \apj, 525, 492

\bibitem[{Gibbons(1972)}]{Gibbons:1972uw}
Gibbons, J.~F. 1972, in Proceedings of the IEEE

\bibitem[{Guhathakurta \& Draine(1989)}]{1989ApJ...345..230G}
Guhathakurta, P., \& Draine, B.~T. 1989, \apj, 345, 230

\bibitem[{Hoang {et~al.}(2015)Hoang, Lazarian, \&
  Schlickeiser}]{2015ApJ...806..255H}
Hoang, T., Lazarian, A., \& Schlickeiser, R. 2015, \apj, 806, 255

\bibitem[{Itoh {et~al.}(2009)Itoh, Duffy, Khakshouri, \&
  Stoneham}]{2009JPCM...21U4205I}
Itoh, N., Duffy, D.~M., Khakshouri, S., \& Stoneham, A.~M. 2009, Journal of
  Physics: Condensed Matter, 21, 4205

\bibitem[{Jenkins(2009)}]{2009ApJ...700.1299J}
Jenkins, E.~B. 2009, The Astrophysical Journal, 700, 1299

\bibitem[{Johnson \& Brown(1982)}]{1982NIMPR.198..103J}
Johnson, R.~E., \& Brown, W.~L. 1982, Nuclear Instruments and Methods In
  Physics Research, 198, 103

\bibitem[{Leger {et~al.}(1985)Leger, Jura, \& Omont}]{1985A&A...144..147L}
Leger, A., Jura, M., \& Omont, A. 1985, A\&A, 144, 147

\bibitem[{Linsky \& Wood(1996)}]{1996ApJ...463..254L}
Linsky, J.~L., \& Wood, B.~E. 1996, Astrophysical Journal, 463, 254

\bibitem[{Liu {et~al.}(2001)Liu, Neumann, Trautmann, \&
  M{\"u}ller}]{2001PhRvB..64r4115L}
Liu, J., Neumann, R., Trautmann, C., \& M{\"u}ller, C. 2001, Physical Review B
  (Condensed Matter and Materials Physics), 64, 184115

\bibitem[{Lubin(2016)}]{2016arXiv160401356L}
Lubin, P. 2016, JBIS, 69, 20

\bibitem[{Mathis {et~al.}(1983)Mathis, Mezger, \&
  Panagia}]{1983A&A...128..212M}
Mathis, J.~S., Mezger, P.~G., \& Panagia, N. 1983, A\&A, 128, 212

\bibitem[{Meftah {et~al.}(1994)Meftah, Brisard, Costantini, Dooryhee, Hage-Ali,
  Hervieu, Stoquert, Studer, \& Toulemonde}]{1994PhRvB..4912457M}
Meftah, A., Brisard, F., Costantini, J.~M., {et~al.} 1994, Physical Review B,
  49, 74839

\bibitem[{Musci {et~al.}(2012)Musci, Weryk, Brown, Campbell-Brown, \&
  Wiegert}]{2012ApJ...745..161M}
Musci, R., Weryk, R.~J., Brown, P., Campbell-Brown, M.~D., \& Wiegert, P.~A.
  2012, \apj, 745, 161

\bibitem[{Poppe(2016)}]{Poppe:2016gr}
Poppe, A.~R. 2016, Icarus, 264, 369

\bibitem[{Seitz(1949)}]{Seitz:1949fe}
Seitz, F. 1949, Discussions of the Faraday Society, 5, 271

\bibitem[{Silk \& Barnes(1959)}]{1959PMag....4..970S}
Silk, E. C.~H., \& Barnes, R.~S. 1959, Philosophical Magazine, 4, 970

\bibitem[{Szenes(1997)}]{1997NIMPB.122..530S}
Szenes, G. 1997, Nuclear Inst. and Methods in Physics Research, 122, 530

\bibitem[{Tielens {et~al.}(1994)Tielens, McKee, Seab, \&
  Hollenbach}]{1994ApJ...431..321T}
Tielens, A. G. G.~M., McKee, C.~F., Seab, C.~G., \& Hollenbach, D.~J. 1994,
  \apj, 431, 321

\bibitem[{Tombrello(1994)}]{1994NIMPB..94..424T}
Tombrello, T.~A. 1994, \nimpb, 94, 424

\bibitem[{Toulemonde {et~al.}(2000)Toulemonde, Dufour, Meftah, \&
  Paumier}]{2000NIMPB.166..903T}
Toulemonde, M., Dufour, C., Meftah, A., \& Paumier, E. 2000, Nuclear
  Instruments and Methods in Physics Research Section B, 166, 903

\bibitem[{Toulemonde {et~al.}(2004)Toulemonde, Trautmann, Balanzat, Hjort, \&
  Weidinger}]{2004NIMPB.216....1T}
Toulemonde, M., Trautmann, C., Balanzat, E., Hjort, K., \& Weidinger, A. 2004,
  Nuclear Instruments and Methods in Physics Research Section B, 216, 1

\bibitem[{Wang {et~al.}(1994)Wang, Dufour, Paumier, \&
  Toulemonde}]{1994JPCM....6.6733W}
Wang, Z.~G., Dufour, C., Paumier, E., \& Toulemonde, M. 1994, Journal of
  Physics: Condensed Matter, 6, 6733

\bibitem[{Weingartner \& Draine(2001{\natexlab{a}})}]{2001ApJ...548..296W}
Weingartner, J.~C., \& Draine, B.~T. 2001{\natexlab{a}}, \apj, 548, 296

\bibitem[{Weingartner \& Draine(2001{\natexlab{b}})}]{2001ApJS..134..263W}
Weingartner, J.~C., \& Draine, B.~T. 2001{\natexlab{b}}, \apjs, 134, 263

\bibitem[{Ziegler(1999)}]{1999JAP....85.1249Z}
Ziegler, J.~F. 1999, JAP, 85, 1249

\bibitem[{Ziegler {et~al.}(2010)Ziegler, Ziegler, \&
  Biersack}]{2010NIMPB.268.1818Z}
Ziegler, J.~F., Ziegler, M.~D., \& Biersack, J.~P. 2010, \nimpb, 268, 1818

\end{thebibliography}
\end{document}